\documentclass[parskip]{scrartcl}

\usepackage[margin=15mm]{geometry}

\usepackage[margin=15mm]{geometry}
\usepackage{tikz}
\tikzstyle arrowstyle=[scale=1]
\tikzstyle directed=[postaction={decorate,decoration={markings,
    mark=at position .65 with {\arrow[arrowstyle]{stealth}}}}]
\tikzstyle reverse directed=[postaction={decorate,decoration={markings,
    mark=at position .65 with {\arrowreversed[arrowstyle]{stealth};}}}]

\usepackage{tikz} \usetikzlibrary{arrows,shapes,positioning}
\usetikzlibrary{decorations.markings}
\tikzstyle arrowstyle=[scale=1.6]
\tikzstyle directed=[postaction={decorate,decoration={markings,
    mark=at position .65 with {\arrow[arrowstyle]{stealth}}}}]
\tikzstyle reverse directed=[postaction={decorate,decoration={markings,
    mark=at position .65 with {\arrowreversed[arrowstyle]{stealth};}}}]

\usepackage[hang,small,bf]{caption}

\usepackage{wrapfig}

\usepackage{color,epsfig,amsfonts,amssymb}
\newcommand{\comma}{\, , \; \; }
\newcommand{\period}{\, .}
\newcommand{\eq}{\; = \;}
\newcommand{\sep}{\, , \;\;}
\newcommand{\spce}{\;\;\;\;\;\;}
\newcommand{\negspce}{\! \! \! \! \! \! \! \! \! \! \!  }
\newcommand{\be}{\begin{equation}}
\newcommand{\bd}{\begin{displaymath}}
\newcommand{\ee}{\end{equation}}
\newcommand{\ed}{\end{displaymath}}
\newcommand{\ba}{\begin{eqnarray}}
\newcommand{\ea}{\end{eqnarray}}
\newcommand{\trace}{{\mathrm{trace}} \; }

\newcommand{\minus}{\! - \!}

\newcommand{\half}{{\textstyle \frac{1}{2}}}

\newcommand{\hV}{{\widehat V}}

\newcommand{\mybf}{\normalfont \bfseries}
\newcommand{\myit}{\normalfont \itshape}
\newcommand{\myrm}{\mathrm}

\newcommand{\sn}{\, \mathrm{sn} \, }
\newcommand{\cn}{\, \mathrm{cn} \,  }
\newcommand{\dn}{\, \mathrm{dn} \, }
\newcommand{\vb}{\overline{v}  }

\newcommand{\kay}{k}

\newcommand{\itt}{\normalfont \itshape}

\newcommand{\bff}{\normalfont \bfseries}

\newcommand{\e}{{\mathrm{e}}}

\renewcommand{\theequation}{\arabic{section}.\arabic{equation}}

\title{The bulk, surface and corner free energies of the square lattice Ising model }

\author{ R.J. Baxter\\
{\protect \small  Mathematical
Sciences Institute,  The Australian National}\\
{\protect  \small  University,
 Canberra, A.C.T. 0200, Australia }}

\date{\small 17 July  2016}

\begin{document}


\maketitle

\abstract{ We use Kaufman's spinor method to calculate the bulk, surface and 
 corner  free energies 
$f_b, f_s, f_s', f_c$  of the anisotropic square lattice zero-field Ising model
for the  ordered ferromagnetic case. For $f_b, f_s, f'_s$ our results of course
agree with the early work of Onsager, McCoy and Wu. We also find 
agreement with the conjectures made by  Vernier and Jacobsen (VJ)  for the isotropic case. 
We note that the corner free energy $f_c$  depends only on the elliptic
modulus $k$ that enters the working, and not on the argument $v$, which means that 
VJ's conjecture applies for the full anisotropic model. 
The only aspect of this paper that is new is the actual derivation of $f_c$, but by 
reporting all four free energies together we can see interesting structures linking them.}



 \vspace{5mm}

{ \mybf{ KEY WORDS: } } Statistical mechanics, lattice models, exactly solved models,
surface and corner free energies
 \tableofcontents

 \section{Introduction}

Vernier and Jacobsen{\color{blue}\cite{VJ2012}} considered a number of two-dimensional 
lattice models in statistical mechanics that are ``exactly solved" in the sense that their
bulk free energies (and where appropriate their order parameters) have been calculated
exactly. For most of them their surface free energies have not been calculated, and for none
 of them their corner free energies.
 
 If $Z$ is the partition function of a square lattice of $M$ rows and $N$ columns, then we 
 expect on physical grounds that when $M$ and $N$ are large, taking $\beta = 1/k_B T$
 where $k_B$ is Boltzmann's constant and $T$ the temperature, that for a 
 non-critical system ($T \neq T_c$)
 \be \label{fullZ}
 \beta^{-1} \log Z \eq - MNf_b - Mf_s - N f'_s - f_c  +
 \mathrm{O}(\e^{-\gamma M}, \e^{ -\gamma' N }) \comma \ee
 where $f_b, f_s, f'_s$ and $f_c$ are the bulk, vertical surface, horizontal surface and corner
 free energies, respectively, and $\gamma, \gamma'$ each have positive real part.
 

 The ``partition function per site" $\kappa$ is
 \bd \kappa \eq  \e^{-\beta f_b} \eq \lim_{M, N \rightarrow \infty}  Z^{1/MN} \ed
 and for solvable models can usually be written as  a product over $m$ of factors such as 
 $1\pm w q^m$or   $(1\pm w  q^m)^m$, where
 $q, z$ are parameters that occur naturally in the mathematical calculation.
 Vernier and Jacobsen considered only the isotropic cases of these models, when $z$ is
 some power of $q$, and obtained quite long series expansions for 
 $\ \e^{-\beta f_s} = \e^{-\beta f'_s} $ and $e^{-\beta f_c} $. They looked for, and usually found,
 a simple repeat pattern in the product expansion. This enabled them to conjecture results for
 the surface and corner free energies.
 
 Baxter and Owczarek{\color{blue}\cite{OB1989}} calculated the surface free energy of the 
 square lattice Potts model when the number of states per spin  is $Q$ and $Q \leq 4$, when 
 the  system is critical and the free energies are integrals rather than products. Vernier and 
 Jacobsen considered the case  $Q>4$. In another paper{{\color{blue}\cite{RJB2016}} }
  the author has extended his and Owczarek's working to $Q>4$ , thereby obtaining 
  $f_s, f'_s$ exactly. The result does indeed 
 agree with Vernier and  Jacobsen's conjecture for the isotropic case.
 
 In this paper we calculate the bulk, surface and corner free energies of the two-dimensional 
 Ising model on the square lattice for the ferromagnetically ordered case, when $T <T_c$
 
The square lattice Ising model was the first of such models to be solved, the bulk free 
energy  $f_b$ being calculated by Onsager in 1944. {\color{blue}\cite{Onsager1944}} It is 
simpler than other solvable models because the partition function can be written as a 
pfaffian (the square root of an anti-symmetric 
determinant).{\color{blue}\cite{MPW63}} In 1967 McCoy and 
Wu{\color{blue}\cite[eqn.4.24b]{McCoyWu67}\cite[p.126, eqn.4.24b]{MCWbook}} calculated 
the surface free energies $f_s, f_s'$.
Our results for these quantities do indeed agree with theirs, as we show in Appendix A. 
 

  The partition function of the Ising model is defined in (\ref{partfn}) below and contains two 
 parameters, the vertical and horizontal interaction coefficients $H$ and $H'$. In terms of the 
 elliptic functions that we introduce, these correspond to $\kay$ and $v$ 
 (the modulus and an argument),  or equivalently  to the $q$ and $w$ defined in
  (\ref{defqw}). 
  
  For the isotropic case
 $H = H'$, so $v$ and $w$ are fixed, as in (\ref{iso}) ($v= iK'/2$, $w= q^{1/4}$).
 Our results for $f_b, f_s, f_s', f_c$ do indeed agree with Vernier and Jacobsen's 
conjectures.

It is known that there is a simple ``inversion relation" method which can usually be used 
to   obtain the bulk free energy of a solved model (i.e. one which  satisfies a 
``Yang-Baxter" relation, which means that the free energies have simple  analyticity 
properties). Here we show in section \ref{sec-inversion}  that  this method can be 
extended to obtain $f_s$ and 
 $f_s'$, and to show that $f_c$ is independent of the anisotropy parameter $v$ (or $w$). 
Unlike the main derivation in this paper, this inversion relation method makes some 
assumptions, notably that $e^{-\beta f_b},  e^{-\beta f_s}, e^{-\beta f_s'}, e^{-\beta f_c}$ 
are analytic functions of $w$ in an annulus containing the ``inversion points" $w^2 =1$ 
and $w^2=q$, except possibly for known poles or zeros at those points. It is therefore not 
rigorous, but it provides a much easier route to the calculation of the free energies.

The self-dual Potts  model has very similar properties.{{\color{blue}\cite{RJB2016}} }
For that and other  solvable models O'Brien, Pearce, 
Behrend and Batchelor{\color{blue}\cite{Pearce95, Batchelor96,Pearce97}} 
 have obtained surface free energies by 
using the reflection Yang-Baxter relations together with the inversion identities and 
inversion relations. Much of their work concerns a lattice which differs from ours by 
rotation though $45^{\circ}$, for which the boundaries are changed and one would expect
from simple, plausible but not rigorous, arguments there to be two inversion relations.
However,  Pearce{{\color{blue}\cite[eqn. 52]{Pearcenotes}} } has been able to rotate their
exact results through $45^{\circ}$ to obtain an extra relation for the
surface free energy $f_s'$ of the self-dual Potts model with the same orientation of the 
lattice as here. Possibly these methods could be used to provide a rigorous justification 
of the inversion relation method of section \ref{sec-inversion}.
 
It is also true for the self-dual Potts model that $f_c$ is independent of the anisotropy 
parameter. Thus for the rectangular lattice  $f_c$ is like the order parameters, notably the 
spontaneous magnetization ${\cal M}_0$, which also are independent of the anisotropy  
parameter. (However, the same is probably not true for other lattices, such as the
 triangular.)


Our derivation does manifest an interesting property of the four free energies
 $f_b, f_s, f_s', f_c$. Apart {from} additive terms that are logarithms of simple rational
functions of the Boltzmann weights, they are integrals or sums of  functions that can be 
written as  $A_1 B_1, A_1 B_2, A_2 B_1,  A_2 B_2$, respectively. Thus if one knows  
$f_b, f_s, f_s'$, then
 $f_c$ is determined, to within such an additive term. The additive terms do not affect
 the singularities at the critical point, so one should be able to obtain the critical behaviour 
 of $f_c$ from  a knowledge of $f_b, f_s, f_s'$.  We do note in section \ref{sec-critical} 
  that the free 
 energies have critical singularities of the form $(T_c-T)^{2-\alpha} \log(T_c-T)$, where
 $\alpha = 0,1,1,2$ for  $f_b, f_s, f_s', f_c$, respectively. Identifying the four exponents 
 in an obvious way,  this  implies that 
\be  \alpha_f+\alpha_c = \alpha_s+\alpha_s' \comma \ee
and maybe this equation has more general application.

In this paper we consider only the ferromagnetically ordered case of the Ising model, when 
$H$, $H'$ are positive and $T<T_c$. We expect the extension to the disordered case, 
when $T >T_c$, to be calculable in a similar manner.

We emphasize that our results are exact for the non-critical Ising model.
Cardy and Peschel,{\color{blue}\cite{CardyPeschel88}} and Wu and
 Izmailyan{\color{blue}\cite{Wu2015}}  have considered the model
 at criticality, where one can use and compare with  the predictions of conformal
 field theory.
 
 There has also been much work 
on the surface and corner magnetizations of the Ising 
model,{\color{blue}\cite{Abraham95, Davies97}} 
but we do not discuss this here.


\section{The Ising model partition function}
\setcounter{equation}{0}

\setlength{\unitlength}{1.2pt}
\begin{figure}[hbt]
\begin{picture}(420,160) (25,0)

\put (151,-12) { $H'$ }

\put (226,64) { $H$ }

\put (81,-14) { $1$ }
\put (303,-16) { $N$ }

\put (73,-1) { $1$ }
\put (71,127) { $M$ }

\put (334, 65) {\Large  $\cal L$ }
\put (334.5, 65) {\Large  $\cal L$ }

\thicklines

{\color{blue} 
\put(95,0) {\line(1,0) {38}}
\put(95,45) {\line(1,0) {38}}
\put(95,90) {\line(1,0) {38}}
\put(95,135) {\line(1,0) {38}}

\put(140,0) {\line(1,0) {38}}
\put(140,45) {\line(1,0) {38}}
\put(140,90) {\line(1,0) {38}}
\put(140,135) {\line(1,0) {38}}

\put(185,0) {\line(1,0) {38}}
\put(185,45) {\line(1,0) {38}}
\put(185,90) {\line(1,0) {38}}
\put(185,135) {\line(1,0) {38}}

\put(230,0) {\line(1,0) {38}}
\put(230,45) {\line(1,0) {38}}
\put(230,90) {\line(1,0) {38}}
\put(230,135) {\line(1,0) {38}}

\put(275,0) {\line(1,0) {38}}
\put(275,45) {\line(1,0) {38}}
\put(275,90) {\line(1,0) {38}}
\put(275,135) {\line(1,0) {38}}

\put(91,4) {\line(0,1) {38}}
\put(91,49) {\line(0,1) {38}}
\put(91,94) {\line(0,1) {38}}

\put(136,4) {\line(0,1) {38}}
\put(136,49) {\line(0,1) {38}}
\put(136,94) {\line(0,1) {38}}

\put(181,4) {\line(0,1) {38}}
\put(181,49) {\line(0,1) {38}}
\put(181,94) {\line(0,1) {38}}

\put(226,4) {\line(0,1) {38}}
\put(226,49) {\line(0,1) {38}}
\put(226,94) {\line(0,1) {38}}

\put(271,4) {\line(0,1) {38}}
\put(271,49) {\line(0,1) {38}}
\put(271,94) {\line(0,1) {38}}

\put(316,4) {\line(0,1) {38}}
\put(316,49) {\line(0,1) {38}}
\put(316,94) {\line(0,1) {38}}

\multiput(91.4,0.5)(45,0){6}{\circle{7}}
\multiput(91.4,45.5)(45,0){6}{\circle{7}}
\multiput(91.4,90.5)(45,0){6}{\circle{7}}
\multiput(91.4,135.5)(45,0){6}{\circle{7}}
}

{\color{red}  
\multiput(87,-23)(0,3){7} {$.$}
\multiput(87,-23.2)(0,3){7} {$.$}
\multiput(132,-23)(0,3){7} {$.$}
\multiput(132,-23.2)(0,3){7} {$.$}
\multiput(177,-23)(0,3){7} {$.$}
\multiput(177,-23.2)(0,3){7} {$.$}
 \multiput(222,-23)(0,3){7} {$.$}
 \multiput(222,-23.2)(0,3){7} {$.$}
 \multiput(267,-23)(0,3){7} {$.$}
 \multiput(267,-23.2)(0,3){7} {$.$}
 \multiput(312,-23)(0,3){7} {$.$}
 \multiput(312,-23.2)(0,3){7} {$.$}

 \multiput(87,138)(0,3){9} {$.$}
\multiput(87,138.2)(0,3){9} {$.$}
\multiput(132,138)(0,3){9} {$.$}
\multiput(132,138.2)(0,3){9} {$.$}
\multiput(177,138)(0,3){9} {$.$}
\multiput(177,138.2)(0,3){9} {$.$}
\multiput(222,138)(0,3){9} {$.$}
\multiput(222,138.2)(0,3){9} {$.$}
\multiput(267,138)(0,3){9} {$.$}
\multiput(267,138.2)(0,3){9} {$.$}
\multiput(312,138)(0,3){9} {$.$}
\multiput(312,138.2)(0,3){9} {$.$}

\put (223,154) { $J$ }
\put (223.2,154) { $J$ }
\put (223,-23) { $J$ }
\put (223.2,-23) { $J$ }

}

 \end{picture}
\vspace{1.0cm}

  \caption{ The square lattice $\cal L$ (of 4 rows and 6 columns), indicating the vertical 
  and horizontal interaction coefficients  $H, H'$.}

 \label{sqlattice1}
\end{figure}
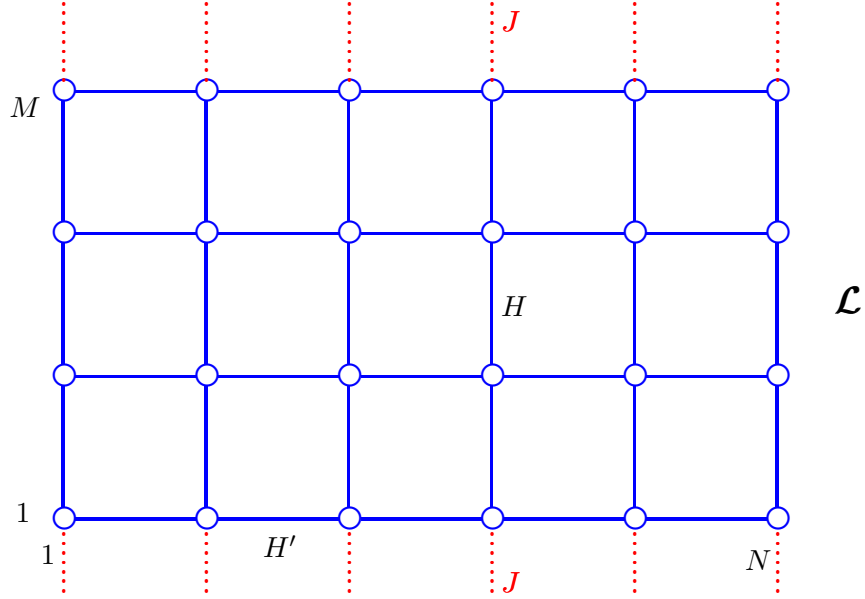

 We consider the Ising model on the square lattice of $M$ rows and $N$ columns, as shown
 by the solid lines and the circles in Fig. \ref{sqlattice1}.  On each site $i$ we place 
 a ``spin" $\sigma_i$, with values
 $-1$ and $+1$. The vertical and horizontal interaction coefficients are $H = \beta J$,
 $H' = \beta J'$ and the partition function is
 \be \label{partfn}
 Z \eq \sum_{\sigma} \exp\left( H \sum_{<ij>}  \sigma_i \sigma_j +H' \sum _{<ij>}
  \sigma_i \sigma_j  \right) \ee
 the first of the two inner sums being over all $(M-1)N$ vertical edges $i,j$, the second
over all $M(N-1)$ horizontal edges $i,j$, and the outer sum over all $2^{MN}$ values of
the $MN$ spins.

Let $\sigma = \{ \sigma_1, \ldots ,\sigma_N \}$ be the spins on a row of the lattice,
and  $\sigma' = \{ \sigma'_1, \ldots ,\sigma'_N \}$ be the spins on the row above. Then
we  can define the $2^N$-dimensional row-to-row transfer matrices $V_1$, $V_2$,
with elements
\bd
\left( V_1 \right)_{\sigma,\sigma'} \eq \prod_{i=1}^N  \e^{H \sigma_i \sigma'_i} \ed

\be \label{defV2}
\left( V_2 \right) _{\sigma,\sigma'} \eq  \exp [ H' \sum_{i=1}^{N-1} \sigma_i \sigma_{i+1} ]
\,   \prod_{i=1}^{N} \delta_{\sigma_i, \sigma'_i} 
 \ee
 The matrix $V_2$ is diagonal.
 
 Let 
\be {\mathbf 1}  \eq  \left(  \begin{array}{cc}
1 & 0 \\
0 & 1 \end{array}  \right) \sep
{\mathbf s} \eq  \left(  \begin{array}{cc}
1 & 0 \\
0 & -1 \end{array}  \right) \sep
{\mathbf c} \eq  \left(  \begin{array}{cc}
0 & 1 \\
1 & 0 \end{array}  \right) \comma \ee
and define $s_j, c_j$ to be the $2^N$-dimensional matrices
\ba s_j \eq  {\mathbf 1}  \otimes  \cdots \otimes {\mathbf 1} \otimes {\mathbf s}
 \otimes {\mathbf 1} \otimes 
\cdots  \otimes {\mathbf 1} \comma \nonumber \\
c_j \eq  {\mathbf 1}  \otimes  \cdots \otimes {\mathbf 1} \otimes {\mathbf c} 
\otimes {\mathbf 1} \otimes 
\cdots  \otimes {\mathbf 1} 
\comma \ea
$\mathbf{s}$, $\mathbf{c}$  on the RHS being in position $j$. Then
\be   \label{defV1}
V_1 \eq (2 \, \sinh 2H)^{N/2} \, \exp \{  H^* \sum_{i=1}^N c_i  \} \ee
\be  \negspce \negspce  V_2 \eq \exp \{ H' \sum_{i=1}^{N-1} s_i s_{i+1}  \}  \comma \ee
where, as in Onsager,{\color{blue}\cite[eq. 14]{Onsager1944}}
\be \label{defH*}
\tanh H^* \eq \e^{-2H} \period \ee


 \subsection{Spinor representatives}
 Kaufman{\color{blue}\cite{Kaufman1949}} simplified Onsager's calculation by using
 spinor (or free-fermion) operators, where one represents the $2^N$-dimensional matrices
 $V_1, V_2$ by $2N$-dimensional matrices  $\hV_1, \hV_2$. Here we use this method.
 
 For $j = 1, \ldots ,N$, define 
 \be \Gamma_{j} \eq  c_1 c_2 \cdots c_{j-1}s_j \sep 
 \Gamma_{j+N} \eq  i \, \Gamma_{j}  \, c_j \ee
 and note that
 \be \label{quad}
 c_j = -i \, \Gamma_{j} \Gamma_{j+N} \sep s_j s_{j+1}  = -i \, 
 \Gamma_{j+N} \Gamma_{j+1}  \period \ee
 Then 
  \be \label{anticomm}
  \Gamma_i \Gamma_j +  \Gamma_j \Gamma_i  \eq 2 \, \delta_{i j} \, I  \comma \ee
  $I$ being the $2^N$-dimensional identity matrix. Let $X,Y$ be two matrices such that
  \be X  \,  \Gamma_i  \,  X^{-1} =  \sum_{j=1}^{2N} x_{j,i} \, \Gamma_j \sep \spce 
  Y \, \Gamma_i \, Y^{-1} = \sum_{j=1}^{2N} y_{j,i} \, \Gamma_j  \ee
and set $T = XY$. Then,  for $i = 1, \ldots 2N$,
  \be  T\, \Gamma_i \, T^{-1} =  \sum_{j=1}^{2N} t_{ji} \, \Gamma_j \comma \ee
  where 
  \be t_{ij} \eq \sum_{m=1}^{2N}  x_{im} \, y_{mj} \period \ee
  It follows that such matrices form a group $\cal G$. Let $\widehat{X} $ be the 
  $2N$-dimensional 
  matrix with elements  $x_{ij}$, and similarly for $\widehat{Y} $, $\widehat{T} $,
  then we call $\widehat{X}, \widehat{Y}, \widehat{T} $ the {\em representatives}
  of $X,Y,T$ and $T = XY$ implies  $\widehat{T} = \widehat{X} \widehat{Y}$.
   
For arbitrary $\rho_1, \ldots \rho_N$,  let
 \be X \eq \exp \{  \sum_{j=1}^{N} \rho_j \, c_j  \} \eq \exp \{ -i \, 
  \sum_{j=1}^{N}  \rho_j \, \Gamma_{j} \Gamma_{j+N}  \}  \comma \ee
 then, using (\ref{anticomm}),
 \bd X \, \Gamma_{j} X^{-1} \eq \cosh 2 \rho_j  \, \Gamma_{j} + i \sinh 2 \rho_j  \, 
 \Gamma_{j+N} \ed
  \bd X \, \Gamma_{j+N} X^{-1} \eq - i \sinh 2 \rho_j  \, \Gamma_{j}  +  \cosh 2 \rho_j  \, 
  \Gamma_{j+N} \ed
  so 
  \be \label{form}
  \widehat{X} \eq  \left( \begin{array}{cc}
A & -i \, B  \\
i \, B^T & C  \end{array} \right) \ee
 where the $A, B, C $ are $N$-dimensional diagonal matrices
 with entries
 \be A_{ij} \eq C_{ij} \eq ( \cosh 2 \rho_j ) \; \delta_{ij} \sep B_{ij} \eq  (\sinh 2 \rho_j ) \; \delta_{ij} 
\ee

  The $2N$ eigenvalues of $\widehat{X}$ are therefore  $\e^{2 \rho_j}$ and  $\e^{-2 \rho_j}$ 
  for $j = 1 , \ldots N$. 
  
  For {\em any} matrix $X$ within $\cal G$, it follows that if the eigenvalues of its
  representative $\widehat{X}$ are 
  $\e^{2 \rho_j}$ and  $\e^{-2 \rho_j}$ (for $j = 1, \ldots N$)  then there must be an 
  invertible matrix $P$ (also within the group) such that 
  \be P \, X \, P^{-1}  \eq R  \, \exp \{  \sum_{i=1}^{N} \rho_i c_i  \}  \period \ee
Thus the eigenvalues of $X$ are 
  \be  R \, \e^{\pm \rho_1 \pm \rho_2  \pm \cdots \pm \rho_N } \ee
  for all $2^N$ choices of the signs, and the scalar factor $R$ can be determined from
  \be \label{defR}
   {2^N} \, \log R  \eq  
  \log \det X   \eq  \trace \log X  \period \ee
    The trace of $X$ is therefore
  \be \label{trX}
  \trace X \eq R \prod_{j=1}^N 2 \cosh \rho_j  \eq R \prod_{j=1}^N (\Lambda_j +2+ 
  1  /\Lambda_j)^{1/2} \ee
  where the $\Lambda_j = \e^{2 \rho_j}$ and  the eigenvalues of $\widehat{X}$
  are $\Lambda_1, \ldots \Lambda_N, \Lambda_1^{-1}, \ldots \Lambda_N^{-1}$.
  

{From} (\ref{quad}), the  matrices $V_1, V_2$ are exponentials of quadratic forms 
in the $\Gamma_j$:
\be V_1 \eq (2 \, \sinh 2H)^{N/2} \, \exp \{  - i \, H^* \sum_{i=1}^N \Gamma_{j} 
\Gamma_{j+N} \} \ee
\be V_2 \eq  \exp \{  - i \, H' \sum_{i=1}^{N-1} \Gamma_{j+N} \Gamma_{j+1} \} \period \ee
Then, for $ 1\leq j \leq N$,
\bd V_1 \, \Gamma_{j} V_1^{-1} \eq \cosh 2 H^*  \, \Gamma_{j} + i \sinh 2 H^* \, 
\Gamma_{j+N} \ed
\bd V_1 \, \Gamma_{j+N} V_1^{-1} \eq - i \sinh 2H^* \, \Gamma_{j}  +  \cosh 2 H^* \, 
\Gamma_{j+N} \ed

\noindent and for for $ 1\leq j \leq N-1$,
\bd V_2 \, \Gamma_{j+N} V_2^{-1} \eq \cosh 2 H'  \, \Gamma_{j+N} + i \sinh 2 H'
  \, \Gamma_{j+1} \ed
\bd V_2 \, \Gamma_{j+1} V_2^{-1} \eq - i \sinh 2H' \, \Gamma_{j+N}  +  \cosh 2 H' 
 \, \Gamma_{j+1} \period \ed
 Since $V_2$ commutes with $\Gamma_1$ and $\Gamma_{2N}$, 
 \bd V_2 \, \Gamma_{1} V_2^{-1} = \Gamma_{1}  \sep
 V_2 \, \Gamma_{2N} V_2^{-1} = \Gamma_{2N} \period \ed
 
 It follows that $V_1, V_2$, belong to $\cal G$ and have representatives
$\hV_1, \hV_2$ of the form (\ref{form}). Set
 \be c^* = \cosh 2 H^* \sep s^* = \sinh 2 H^*  \sep c' = \cosh 2 H' \sep s' = 
 \sinh 2 H' \comma  \ee
 then for  $V_1$, 
 \be A =  C  = c^{*} \,  \mathbf{1} \sep B =  s^{*} \,  \mathbf{1}  \ee
 where $\mathbf{1}$ is now the identity $N$ by $N$ matrix. For
 $V_2$,
 \bd A \eq \left( \begin{array}{ccccc}
1 & 0 & .. & 0 & 0   \\
0 & c' & .. & 0 & 0   \\
.. & .. & .. & .. & ..   \\
0 & 0 & ..  & c' & 0   \\
0 & 0 & ..  & 0 & c'     \end{array} \right) \sep 
C \eq \left( \begin{array}{ccccc}
c' & 0 & .. & 0 & 0   \\
0 & c' & .. & 0 & 0   \\
.. & .. & .. & .. & ..   \\
0 & 0 & ..  & c' & 0   \\
0 & 0 & ..  & 0 & 1    \end{array} \right) \comma \ed
 \be B \eq \left( \begin{array}{cccccc}
0 & 0 & .. & 0 & 0 & 0   \\
-s' & 0 & .. & 0 & 0 & 0    \\
0 & -s' & .. & 0  & 0 & 0   \\
.. & .. & .. & ..  & .. & ..   \\
0 & 0 & ..  & -s' & 0 & 0   \\
0 & 0 & ..  & 0 & -s' & 0     \end{array} \right) \ee
so for $V_2$ the $N$ by $N$ matrices
$A, C$ are diagonal and $B$ is one-off diagonal.


\subsection{The top-to-bottom boundary condition}
We handle the top and bottom boundary conditions as follows.
We link the top and bottom rows by extra vertical edges, shown as dotted (red) lines
in Figure \ref{sqlattice1}, and allow the top and bottom spins to interact with an
 interaction coefficient $J$. This changes the boundary conditions to the familiar 
 cylindrical ones,  but we can readily regain the original open boundary conditions by 
 taking the limit $J \rightarrow 0$.
Let $W$  be the transfer matrix of this row, given by $V_1$, but with $H$ replaced
by $J$. Then the partition function is
\be \label{Zastrace}
Z \eq {\mathrm{trace}} ( W V_2 V_1 V_2 \cdots V_1 V_2 ) \comma \ee
there being $M$ factors $V_2$, and $M-1$ factors $V_1$, in the matrix product.
We note that 
\be W V_2 V_1 V_2 \cdots V_1 V_2  \eq F \times {\mathrm{ product \; of \;  
traceless  \; matrices} } \ee
where
\be \label{defF}
F \eq (2 \, \sinh 2J)^{N/2} \, (2 \, \sinh 2H)^{N(M-1) /2}  \period \ee

The matrix $W$, like $V_1$ and $V_2$, belongs to the group $\cal G$. Let
\be \label{defU}
 \widehat{U} \eq \hV_2 \hV_1 \hV_2 \cdots \hV_2 \ee
Then the representative of $W V_2 V_1 V_2 \cdots V_1 V_2  $ is
\bd \widehat{W}  \, \widehat{U} \period  \ed

From (\ref{defR}) and (\ref{trX}), it follows that
\be \label{behZ}
Z \eq F \,  \prod_{j=1}^N (\Lambda_j +2+ 
  1  /\Lambda_j)^{1/2}  \comma \ee
  where $\Lambda_1, \ldots \Lambda_{N},\Lambda_1^{-1}, \ldots \Lambda_{N}^{-1} $ are 
  the $2N$ eigenvalues of  $ \widehat{W}  \, \widehat{U} $.

Now we take the limit $J \rightarrow 0$ to regain the desired closed boundary
conditions. Analogously to (\ref{defH*}),
\be \tanh J^* = \e^{-2J} \comma \ee
so $J^* \rightarrow \infty $. Then  all the non-zero elements of $ \widehat{W} 
\rightarrow \infty$ and 
\be \lim_{J \rightarrow 0} \,  2 J \, \widehat{W}  = P =  \left( \begin{array}{cc}
I  & -i \, I  \\  i \, I  & I   \end{array} \right) \period \ee
This matrix $P$ is of rank $N$. It follows that  we can choose the $\Lambda_1, \ldots
, \Lambda_N$  to be of order $1/J$, tending to $\infty$, while the other $N$ eigenvalues of 
$ \widehat{W}  \, \widehat{U} $ are of order $J$, tending to zero.

Hence (\ref{behZ}) becomes
\be Z \eq F \,  (\Lambda_1 \cdots \Lambda_N )^{1/2}  \ee
and now $2 J \Lambda_1, \ldots , 2 J \Lambda_N$ are the non-zero eigenvalues
of   $P  \, \widehat{U}$, which in turn are the eigenvalues of the $N$ by $N$ matrix 
\be \label{defQ}
Q \eq (I \;  -i \, I ) \; \hV_2 \hV_1 \hV_2 \cdots  \hV_2 \; 
\left( \begin{array}{c}
I \\  i \, I   \end{array} \right) \period \ee

{From} (\ref{defF}) and (\ref{defU}), it follows that
\be \label{ZdetQ}
Z \eq 2^{N/2} (2 \sinh 2 H)^{N(M-1)/2} \;   ( \det Q )^{1/2} \ee
so we have reduced the problem to one  of calculating an $N$ by $N$ determinant.


\section{Calculation of $\det Q$}
\label{calcdetQ}
\setcounter{equation}{0}

Eqn. (\ref{defQ}) can be written
\be \label{defQ1}
Q \eq (I \;  -i \, I ) \; (\hV_2 \hV_1)^{M-1} \hV_2  \; 
\left( \begin{array}{c}
I \\  i \, I   \end{array} \right) \comma \ee
which leads us to  look for the eigenvalues $\lambda$  of $\hV_2 \hV_1$. Let 
$\mathbf y$ be one such  eigenvector, and $\mathbf x$ a
vector related to it by the eigenvalue equations
\be \label{eig}
\lambda \,  {\mathbf y}  = \hV_1 \, {\mathbf x}   \sep  {\mathbf x}  =  \hV_2  \,  
{\mathbf y }  \period \ee
These $\mathbf{x},\mathbf{y}$ are of dimension $2N$, we can write the equations more 
explicitly if we define $N$-dimensional vectors $x,x',y,y'$ so that
\be \label{defXY}
{\mathbf x } = \left( \begin{array}{c}
x \\  x'   \end{array} \right) \sep  {\mathbf y } = \left( \begin{array}{c}
y \\  y'   \end{array} \right)   \ee


Then the eigenvalue  equations  (\ref{eig}) are
\bd 
\lambda \, y_j  = c^* \, x_j  - i s^* \, x'_{j} \sep
\lambda \, y'_j =  i \, s^* \, x_j +  c^* \, x'_{j}  \comma  \ed
\be \label{yj} x_j  =  c' \, y_j  + i s' \, y'_{j \! - \! 1}\sep 
x'_{j}  =  - i s' \, y_{j \! + \! 1}   + c' \, y'_{j}  \comma  \ee
where $j = 1, \ldots , N$, except that $j=1$ is excluded in the third equation, and $j = N$
in the fourth. Instead, the corresponding equations are 
\be \label{bdy}
 x_1 = y_1 \sep x'_N = y'_N \period \ee


We first look for a solution of the form
\ba x_j   =  A z^{j-1} &, &  y_j   =  B z^{j-1} \nonumber \\
 x'_j   =  A' z^{j-1} &, &  y'_j   =  B' z^{j-1} \ea
 and find it works for the equations  (\ref{yj}) provided that 
 \bd \lambda B =  c^* A - i s^* A' \sep  \lambda B'  =   i s^* A + c^* A' \ed
  \be \label{eqnsAB}
A =  c' B + i s' B' /z \sep
A' =  -  i s'  z B + c' B' \period \ee
 These are four homogeneous linear equations in $A, A', B, B'$. The determinant must 
 be zero, which gives
 \be \label{deflambda}
 \lambda^2  - [ 2 c' c^* -s' s^* (z+z^{-1} ) ] \lambda  +1 = 0  \period \ee
 This equation is unchanged by replacing $z$ by $z^{-1}$, so if (\ref{yj})
 is one possible ansatz, another one for the same value of the eigenvalue 
 $\lambda$ is obtained by inverting $z$. 
 
 The eqns. (\ref{eqnsAB}) are unchanged by inverting $z$ and simultaneously 
 replacing $A, A', B, B'$ by $-A',A,-B',B$; also, if
$\lambda$ is fixed, the equations (\ref{yj}) are linear. The more  general
 ansatz 
 \ba \label{ans}
 x_j   =  A z^{j-1} - g A' z^{1-j} &, 
&  y_j   =  B z^{j-1} - g B'  z^{1-j} \nonumber \\
 x'_j  =  A' z^{j-1} + g  A z^{1-j} &, &  y'_j    =  B' z^{j-1} +  g B z^{1-j} 
\ea
(for $j = 1, \ldots, N$)  therefore satisfies equations (\ref{yj}) for arbitrary $g$, 
provided  only that  (\ref{eqnsAB}) 
holds.  

We introduce a parameter $\alpha$ such that
\be g =  i z^{N-1} \alpha \period \ee
Then (\ref{ans}) becomes
\ba x_j   =  A z^{j-1} - i \alpha A' z^{N-j} &, 
&  y_j   =  B z^{j-1} - i \alpha B'  z^{N-j} \nonumber \\
 x'_j  =  A' z^{j-1} + i \alpha  A z^{N-j} &, &  y'_j    =  B' z^{j-1}  +  i  \alpha B z^{N-j} 
\ea

We now attempt to satisfy the boundary conditions (\ref{bdy}). It is
convenient to define $y'_0$ by the third of the equations (\ref{yj}) (with $j=1$),
and $y_{N+1}$ by the fourth (with $j=N$). 
Then we can  write the boundary conditions
as
\be (c'-1) y_1 = -i \, s'y'_0  \sep (c'-1) y'_N  = i \, s' y_{N+1}  \comma
\ee

These $y'_0$ and $y_{N+1}$ will also be given by (\ref{ans}), so we obtain
\ba \label{bdyc}
(c'-1)B z  +i \, s' B'   & = & i z^N \alpha  [(c'-1)B'-i s' z B]  \nonumber \\
 (c'-1) B z+ i \, s' B' & = &  i z^{N}  \alpha^{-1} [(c'-1) B' - i s' B z ] \ea
We take
 \be \label{alphasq} \alpha^2 = 1 \comma \ee
 i.e. $\alpha =  +1$ or $-1$, to ensure that these equations are the same.
 That we can do this is a reflection of the fact that $\hV_1$ and $\hV_2$
 are both invariant under replacing rows and columns $1, \ldots 2N$ by 
 $2N, \ldots, 1$, and negating rows and columns $N+1, \ldots, 2N$.


The $c', s'$ are related to one another, as are $c^*, s^*$. We express
them in terms of 
\be \label{defut}
u = \tanh H' \sep t = \tanh H^* = \e^{-2H}  \ee
as
\be \label{exprscs}
c'= \frac{1+u^2}{1-u^2} \sep s' = \frac{2 u}{1-u^2} \sep
 c^*= \frac{1+t^2}{1-t^2} \sep s^*= \frac{2 t}{1-t^2} \period \ee
 Then 
 \be \label{relnstu}
 e^{-2H}  = t \sep e^{-2H' } = \frac{1-u}{1+u} \sep 
 \sinh 2H = \frac{1-t^2}{2t} \period \ee
 
 Here we consider the ordered ferromagnetic phase of the Ising model, when
 $H,H',H^*$ are all positive and
  \be \label{lowtemp}
  \sinh 2H \, \sinh 2H' >1 \period \ee
  Then  $t,u$ are real  and 
  \be  \label{tu}
  0 < t < u < 1 \period  \ee

The eqns. (\ref{bdyc}) both become
\be \label{BpB}
B' = i z  B (u-\alpha \, z^N)/(1-\alpha \, u \, z^N ) \period  \ee

We can eliminate $\lambda, A, A'$ between the equations (\ref{eqnsAB}) to obtain

\bd [(u+u^{-1})z -(t+t^{-1})z^2] B^2 + 2 i (1-z^2) B B' +[  (u+u^{-1})z  - t-t^{-1}] B'^2
= 0 \ed

Making the substitution (\ref{BpB}), we find that $\alpha$ only enters the numerator
via $\alpha^2$. Using (\ref{alphasq} ), we obtain
$P(z) = 0$, where
\ba \label{pol}
P(z)  & = &  z^{2N} (z -tu)(z-u/t)-(1-tuz)(1-uz/t)  \nonumber \\
 & = &  z^{2N+2} -c_1z^{2N+1} + \cdots +c_1 z  - 1  \ea
 where  $c_1 = u (t+1/t)$.

This $P(z)$ is a polynomial in $z$, of degree $2N+2$. Its zeros are
\be   \label{order}  1, -1, z_1, z_2, \ldots , z_N, z_N^{-1}, z_{N-1}^{-1}, \ldots , z_1^{-1} \ee
Substituting $z = 1$ or $z=-1$ into the above equations for ${\mathbf x}, {\mathbf y}$
gives ${\mathbf x} = {\mathbf y } = {\mathbf 0 }$, so these solutions are spurious and we ignore them.
As indicated in Figure.\ref{posnzs}, the remaining $2N$ zeros occur in inverse pairs, 
$2N-2$ of them being 
on the unit circle, and 2 on the positive real axis. We choose $z_1, \ldots z_{N-1}$ to lie on 
the unit circle, in the upper half-plane, ordered from left to right,
 and $z_N$ to lie on the real axis, between $0$ 
and $1$, as indicated in Figure \ref{posnzs}.  The corresponding
eigenvalues $\lambda_1, \ldots , \lambda_N$ are real and positive. If we take
\be \label{alphak}
\alpha_j = (-1)^{N-j} \comma \ee
for all $j$, then 
\be  \lambda_j  > 1 \; {\mathrm{for}} \;\;  j = 1, \ldots , N \;   ;   \;  \; \;  \;  \lambda_j  < 1 
 \; {\mathrm{for}} \;\;  j = N+1, \ldots , 2N  \period \ee
 
Using (\ref{pol}), 
\be   \alpha_j \,  z_j^N \eq \sqrt \frac{(1-tuz_j)(uz_j-t)}{(z_j-tu)(u-tz_j)} \ee
where the square root should be taken to be in the right-half  of the complex plane 
for $1 \leq j \leq N$, in the left-half plane for $N+1 \leq j \leq 2N$. (It is 
straightforward to verify that these choices are consistent in the limit $t \rightarrow 0$. 
By continuity it follows that they are consistent for all $u, t$ satisfying (\ref{tu}).)



\setlength{\unitlength}{1pt}
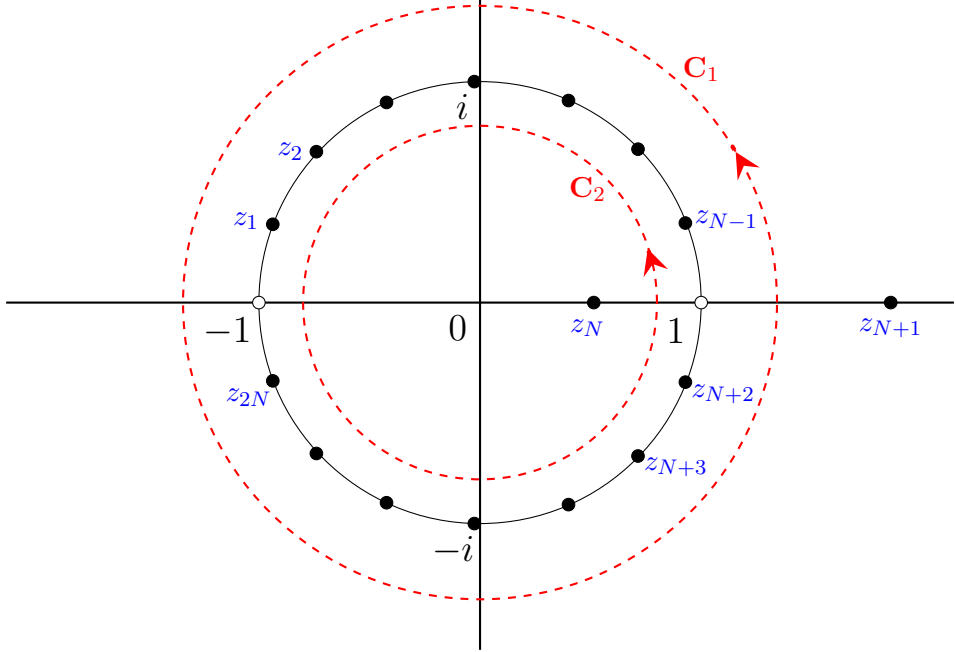
\begin{figure}[hbt]
\begin{picture}(420,240) (-30,0)

\begin{tikzpicture}[scale=0.84]
\draw[thick] (2,-4) -- (17,-4);
\draw[thick] (9.5,-9.5) -- (9.5,1.5);
\draw (9.5,-4) circle (3.5cm);

 \node at (5.50,-4.45) { \Large $-1$};
 \node at (12.6,-4.45) {\Large $1$};
  \node at (9.15,-0.9) { { \mybf \Large$i $}};

   \node at (9.00,-7.92) { { \mybf \Large$-i $}};
  \node at (9.15,-4.4) { \mybf \Large $0$};

 {\color{red} 
 \draw[dashed,thick] (9.5,-4) circle (2.8cm);
 \draw[dashed,thick] (9.5,-4) circle (4.7cm);
 \node at (11.2,-2.2) { $\mathbf C_2$};
 \node at (13.0,-0.3) {$\mathbf C_1$};



\draw [directed,ultra thick,dashed] (12.2,-3.3) arc (15:20:2.8) ;
\draw [directed,ultra thick,dashed] (13.65,-1.8) arc (28:32:4.7) ;

 }

   \draw[black,fill=black] (12.75,-2.74) circle (0.6ex);
    \draw[black,fill=black] (12.75,-5.26) circle (0.6ex);
      \draw[black,fill=black] (6.22,-2.76) circle (0.6ex);
    \draw[black,fill=black] (6.22,-5.24) circle (0.6ex);

  \draw[black,fill=black] (12.0,-1.57) circle (0.6ex);
  \draw[black,fill=black] (12.0,-6.43) circle (0.6ex);
    \draw[black,fill=black] (6.91,-1.61) circle (0.6ex);
  \draw[black,fill=black] (6.91,-6.39) circle (0.6ex);

   \draw[black,fill=black] (10.9,-0.8) circle (0.6ex);
   \draw[black,fill=black] (10.9,-7.2) circle (0.6ex);
    \draw[black,fill=black] (8.02,-0.83) circle (0.6ex);
   \draw[black,fill=black] (8.02,-7.17) circle (0.6ex);

      \draw[black,fill=black] (9.41,-0.5) circle (0.6ex);
    \draw[black,fill=black] (9.41,-7.5) circle (0.6ex);
    
    \draw[black,fill=white] (13,-4) circle (0.6ex);
    \draw[black,fill=white] (6,-4) circle (0.6ex);
    \draw[black,fill=black] (11.3,-4) circle (0.6ex);
    \draw[black,fill=black] (16,-4) circle (0.6ex);
    
 {\color{blue} 
  \node at (5.8,-2.7) {$z_1$};
 \node at (6.5,-1.6) {$z_2$};
 \node at (13.4,-2.7) {$z_{N-1}$};
 \node at (11.2,-4.4) {$z_{N}$};
 \node at (16.0,-4.4) {$z_{N+1}$};
 \node at (13.35,-5.45) {$z_{N+2}$};
  \node at (12.6,-6.6) {$z_{N+3}$};
 \node at (5.84,-5.5) {$z_{2N}$};
 }
 
 \end{tikzpicture}
 \end{picture}

 \vspace{0.5cm}

  \caption{ The positions and ordering of the zeros of (\ref{pol});
  $z_j z_{2N+1-j} =1 $ for all $j$. }

 \label{posnzs}
\end{figure}


 For $N$ large, 
$z_N$ is close to $t/u$ and $z_{N+1}$ is close to $u/t$:
\be  \label{zN} z_N = t/u + \mathrm{ O} (t/u)^{2N}  \sep   z_{N+1}  = u/t + \mathrm{ O} 
(t/u)^{2N} \comma \ee
in agreement with (\ref{pol}).

In particular, $\alpha_N = 1 , \alpha _{N+1} = -1$. For all $j$,
\be \label{condsz}
z_{2N+1-j} = z_j^{-1} \sep \lambda_{2N+1-j} = \lambda_j^{-1} \sep \alpha_{2N+1-j} = -\alpha_j \period \ee

This gives all the distinct $2N$ eigenvectors. If $z= z_m$, we write the vectors 
$\mathbf x$, $\mathbf y$ (as given above), as $\mathbf x_m$, $\mathbf y_m$. Let $X$ be 
the $2N$ by $2N$ matrix with columns  $\mathbf x_1, \ldots  \mathbf x_{2N}$.
Similarly for $Y$, and let ${\cal D}$ be the diagonal matrix with entries
$\lambda_1, \ldots,\lambda_{2N}$. Then (\ref{eig}),  (\ref{defQ1})  become
\bd \hV_1 X \eq Y {\cal D} \sep  \hV_2 Y \eq X \comma \spce  \spce \spce \spce \ed
\be \label{Qform} 
Q \eq (I \;  -i \, I ) \; X {\cal D}^{M-1} Y^{-1} \; 
\left( \begin{array}{c}
I \\  i \, I   \end{array} \right) \period  \ee


\subsection{Explicit expressions}
{From} (\ref{BpB}), we can take 
\be B= 1-\alpha \,  u \,  z^N \sep B' = iz (u- \alpha  z^N) \ee
Then
\ba \label{y} y_j  & = & z^{j-1}-z^{2N+1-j}  -\alpha \, u \, z^N (z^{j-1} -z ^{1-j}) \nonumber \\
 y'_j  & = & i u \, (z^{j}-z^{2N-j} ) - i \, \alpha  \, z^N (z^{j} -z ^{-j}) \ea

Substituting these expressions into the RHS of the second pair of the
equations (\ref{yj}  ),
we find
\ba \label{x}  x_j  & = & z^{j-1}-z^{2N+1-j}  + \alpha \, u \, z^N (z^{j-1} -z ^{1-j}) \nonumber \\
 x'_j  & = & - i u \, (z^{j}-z^{2N-j} ) - i \, \alpha  \, z^N (z^{j} -z ^{-j}) \comma \ea
 so $ x_j ,  x'_j$ have the same forms as $y_j, y'_j$, but with $u$ negated.
 
For the moment, regard $z_1, \ldots , z_N$ as arbitrary, but with the other $z's$ given 
by (\ref{condsz}) and the $\alpha_1, \ldots , \alpha_{2N}$ by (\ref{alphak}), and
consider the determinant of $X$. It will be a Laurent multinomial in $z_1, \ldots z_N$.

Column $m$ of $X$ is the column vector with entries $x_1,\ldots x_N, x'_1, \ldots x'_N$,
as given above with $z = z_m$, $\alpha = \alpha_m$. If two even (or two odd) 
columns have the same value of $z$, then, using (\ref{alphak}), the column vectors 
will be the same and the 
determinant of $X$ will vanish. 

It follows that the multinomial expression for $\det X$ contains factors of the form
$z_j-z_{j+2r}$. Arguing similarly, we show in Appendix B that
\be \det Y \eq  \epsilon_N \, 2^N (1 - u^2)^{N-1} \, \prod_{j=1}^N \frac{(1-z_j^2)^2}{z_j^{2N}}
\prod_{m=j+1}^N (z_j-z_m)^2 (1-z_j z_m)^2 \comma \ee
where 

\be \epsilon_N = 1 \; \;  \myrm{if} \;  N\; \myrm{is\; even},  \; \; \epsilon_N = -i  \; \myrm{if} \; \; 
N \; \myrm{is\; odd} \period \ee

We also find that
\be  \label{yty}
X^T\, X = Y^T \, Y =  {\cal W} S   = S {\cal W} \ee
where \be \label{defWS}
{\cal W} = \left( \begin{array}{ccccc}
\tau_1 & 0 & .. & 0 & 0   \\
0 & \tau_2 & .. & 0 & 0   \\
.. & .. & .. & .. & ..   \\
0 & 0 & ..  & \tau_2 & 0   \\
0 & 0 & ..  & 0 & \tau_1     \end{array} \right) \sep
S = \left( \begin{array}{ccccc}
0 & 0 & .. & 0 & 1  \\
0 & 0 & .. & 1& 0   \\
.. & .. & .. & .. & ..   \\
0 & 1& ..  & 0 & 0   \\
1 & 0 & ..  & 0 & 0     \end{array} \right) \period \ee

Taking square roots, choosing the sign appropriately, it follows that
\be \label{detY}
\det Y = \epsilon_N \; \tau_1 \tau_2 \cdots \tau_N \ee
and that

\bd
Q \eq (I \;  -i \, I ) \; X {\cal D}^{M-1} {\cal W}^{-1} \, S \, Y^T 
\left( \begin{array}{c}
I \\  i \, I   \end{array} \right) \period  \ed

Set \be \label{defLR}
L = (I \;  -i \, I ) \; X  \sep R = S \, Y^T 
\left( \begin{array}{c}
I \\  i \, I   \end{array} \right) \comma \ee
so
\be \label{eqnQ} Q = L \, {\cal D}^{M-1} {\cal W}^{-1} R \period  \ee

{From} (\ref{defXY}) and (\ref{defLR}), the elements of $L, R$ are
\be
L_{j,m} \eq  x_j -i \, x'_j  \sep
 R_{2N+1-m,j} = y_j+i \, y'_j  \comma \ee
 where $x_j,  x'_j, y_j,  y'_j$ are defined by (\ref{x}), (\ref{y}) with
 $z = z_m, \alpha = \alpha_m$.
 
Define the $N$ by 
 $2N$ matrix $\tilde{C}$ by
 \be \tilde{C}_{j,m} \eq  -i (z_m^{j-1}+\alpha_m \, z_m^{N-j})( 1+
 \alpha_m \, u \, z_m^{N}-u z_m -  \alpha_m \, z_m^{N+1})/z_m^N \ee
 Then we find that
 \be \label{LRC} L_{j,m} \eq i z_m^N \, \tilde{C}_{j,m} \sep R_{m,j} \eq -i \, z_m^{-N} \,  
 \tilde{C}_{m,j}\ee
where $z=z_m, \alpha = \alpha_m$ are the values of $z, \alpha$ for the  $m$th  
eigenvector.
  We have used the relations $z_{2N+1-m} = 1/z_m$, $\alpha_{2N+1-m} = -\alpha_m$.
  
  Because ${\cal D, W}$ are diagonal, the factors $i z_m^N, -i z_m^{N}$ in (\ref{LRC}) cancel 
  out of (\ref{eqnQ}), so 
  \be \label{Qeq2}
  Q =  \tilde{C} \, {\cal D}^{M-1} {\cal W}^{-1} \tilde{C}^T  \period  \ee
  
  Factor $\tilde{C}$ into two parts by defining another $N$ by $2N$ matrix $\cal C$ and a 
  $2N$ by $2N$ diagonal  matrix $\cal F$ by
  \be \label{defC}
   {\cal C}_{j,m }  = z_m^{j-1}+\alpha_m \, z_m^{N-j} \ee
   \be \label{defF2}
   {\cal F}_{j,m} = -i \, \delta_{jm}  \, ( 1+\alpha_m \, 
  u \, z_m^{N}-u z_m -  \alpha_m \, z_m^{N+1})/z_m^N \comma  \ee
  then $\tilde{C} = {\cal C\, F }$ and (\ref{Qeq2}) becomes
  \be \label{Qeq3}  Q = {\cal C  F \, {\cal D}}^{M-1} {\cal W}^{-1}  {\cal F \, C}^T  \period  \ee



\section{The limit of $M$ large }
\label{sec-mlarge}
\setcounter{equation}{0}
We emphasize that equations (\ref{ZdetQ}), (\ref{Qeq3}) are {\em exact}, giving the 
partition function of a finite $M$ by $N$ lattice. 

We only need the determinant of $Q$, but because the matrix 
$C$ in (\ref{Qeq3})  is not square (it is $N$ by $2N$), 
we cannot see how to simplify $\det Q$ further for finite $M$.

However,  we are interested in the limit of $M$ large (where we can neglect
terms of relative order $\e^{-\gamma M}$, and $M$ enters  (\ref{Qeq3}  )  only via
the explicit power of the diagonal matrix $\cal D$ therein.

Since $\lambda_1$ is the largest of the eigenvalues $\lambda_m$,
when $M$ is large, the elements of $Q$ will be dominated by terms proportional to 
$\lambda_1^{M-1}$. However, this contribution to $Q$ is of rank 1, so has zero
determinant. To obtain a non-zero determinant, we must retain at least $N$ distinct
eigenvalues from the $\lambda_m$ so as to ensure that the truncated matrix $Q$
is non-singular. The choice that gives the largest contribution to
$\det Q$ is to choose the $N$ largest eigenvalues, i.e. $\lambda_1, \ldots,
\lambda_N$. For $M$ large and $N$ fixed, the terms that are neglected 
will be relatively exponentially small, so we should still obtain the first four 
contributions to $\log Z$, as in (\ref{fullZ}).

Keeping only the first $N$ diagonal elements of ${\cal D}$ is the same as
truncating ${\cal D, C, F, W}$ to the $N$ by $N$ matrices $D, C, F, W$
with elements 
\be \label{defCF}
D_{ij} = {\cal D}_{ij} \sep C_{ij} = {\cal C}_{ij} \sep F_{ij} = {\cal F}_{ij} 
\sep W_{ij} = {\cal W}_{ij} \ee
for $i,j = 1, \ldots N$.

Set \be \label{defDelta}
\Delta= \lambda_1 \lambda_2 \cdots \lambda_N \sep 
 \phi= f_{11} f_{22} \cdots f_{NN} \sep  \widehat{\tau} = 
 \tau_1 \tau_2 \cdots \tau_N \ee
then from (\ref{Qeq3})
\be \det Q \eq  \frac{(\det C)^2 \, \phi^{\, 2} \, \Delta^{M-1} }{\widehat{\tau}} 
\period \ee
Using (\ref{detY}), this becomes
\be \label{Qeq5}
\det Q \eq  \frac{\epsilon_N \, (\det C)^2 \, \phi^{\, 2} \, \Delta^{M-1} }{\det Y} 
\period \ee

Let \be \label{defzeta}
\zeta \eq z_1 z_2 \cdots z_N \comma \ee
then from (\ref{defF2})
\be \label{defphi}
\phi \eq (-i)^N \zeta^{-N} \prod_{m=1}^N  (1+\alpha_m \, 
  u \, z_m^{N}-u z_m -  \alpha_m \, z_m^{N+1} ) \ee

We define also
\be  \label{defphip}
\phi ' \eq (-i)^N \zeta^{-N} \prod_{m=1}^N  (1-\alpha_m \, 
  u \, z_m^{N}-u z_m + \alpha_m \, z_m^{N+1} ) \ee
  and set
  \be \label{defmu}
  \eta \eq \prod_{m=1}^N \frac{1+z_m}{1-z_m} \period \ee
  
  We shall find it convenient to break  the RHS of (\ref{Qeq5}) into three parts and to
  write the equation as
  \be \label{detqEG}
  \det Q \eq  E \, G \Delta^{M-1} \comma \ee
  where
  \be \label{defEG}
  E  \eq  \frac{\epsilon_N \, \phi \, \phi' (\det C)^2}{\eta \det Y} \sep
 G \eq \frac{\eta \phi }{\phi' }  \ee
 
 Set \be z_m = \e^{2 i \theta_m } \sep \alpha_m = \e^{ 2 i a_m } \period \ee
 Then, using (\ref{deflambda}),
  \be \log \lambda_m \eq \cosh^{-1} (c' c^* -s' s^* \cos 2 \theta_m ) \ee
  and 
  \be G \eq \prod_{m=1}^N ( \cot \theta_m  ) \;  \frac{\sin ( a_m +N\theta_m + \theta_m ) +
  u \sin(a_m +N \theta_m - \theta_m )}{\cos ( a_m +N\theta_m +\theta_m) +
  u \cos (a_m +N\theta_m -\theta_m )} \period \ee
  
  $\theta_1, \cdots ,\theta_{N-1}$ are real, between 0 and $\pi /2$, while $ \theta_N$ 
  is positive pure imaginary; $a_m$ is either zero or $\pi /2$. It follows that $\log \lambda_m$ 
  and $G$ are real. In fact they are positive.

 \subsection{Calculation of $E$}
 
 We shall show in Appendices A and B that in the $M$ large limit, $E$ is a product
 of powers of simple rational functions of $u$ and $t$. The working is messy and we have to 
 consider separately the cases when $N$ is even and when $N$ is odd.
 The final result is given immediately below in (\ref{Etot}). The author hopes that 
 someone will find a way to simplify the working.
 
We first show in Appendix B that the determinants of $Y$ and $C$
as products of simple factors that are linear in $z_1, \ldots, z_N$.
From (\ref{resY}), (\ref{detc}), (\ref{tc}),
\be \label{YC2}
 \frac{\det Y} { (\det C)^2} \eq \frac{ {\cal L} (1-u^2)^{N-1} }{\zeta^{2N}}
\, \prod_{1  \leq j < m \leq N}^{ \mathbf {\dagger \dagger }} (z_j-z_m)^2 (1-z_j z_m)^2 \comma \ee
where the superfix $\dagger \dagger$ means that the double product
is over all $j,m$ of opposite parity, i.e.  $ j-m =1$ modulo  2, and
\ba {\cal L} & = &  \prod_{j=1}^{N/2} (1+z_{2j-1})^2 (1-z_{2j})^2 \; \; \myrm{if} 
\; N \; \myrm{is \; even} \comma \nonumber \\ 
& = & \half  \epsilon_N \prod_{j=1}^{(N+1)/2} (1-z_{2j-1}^2)^2 \; \; \myrm{if} 
\;  N  \; \myrm{is \; odd }\period \ea

In Appendix C we go on to calculate the $E$ of (\ref{defEG}).
The fact that the double product is restricted to $j,m$ being of opposite parity is 
significant. Factors with the same parity do occur in $\det Y$ and $\det C$, but cancel out
of the ratio (\ref{YC2}). This means that we only need specific values of 
the polynomials $P_1(z), P_2(z) $ defined in  Appendix C, rather than their derivatives, which 
are  considerably more complicated.

The results for  $N$ even and $N$ odd are given in eqns. (\ref{Eeven}) and (\ref{Eodd}). 
They are identical: both give
\be \label{Etot}
  E \eq \frac{ (1-t)^{2N} (1-ut) \, (1-t^2/u^2)^{3/4} }{2^{N-2} (1-t/u) \,
   (1-u^2 \, t^2)^{1/4} \, (1-t^2)^{1/2}} \; \;  \period \ee
   {From} (\ref{tu}) we see that $E$ is positive real. So therefore is the RHS of 
   (\ref{detqEG}).
   

 \section{The free energies as integrals}
 \setcounter{equation}{0}
 
 Define \be \label{defrho}  \rho (z)  \eq \sqrt \frac{(1-tuz)(uz-t)}{(z-tu)(u-tz)} \ee
 where the square root is chosen to be in the right-half plane. Then
 $\rho(z)$ is analytic in the cut $z$-plane denoted in Figure  \ref{posnzs}, where there
 are branch cuts denoted by solid (red) horizontal lines on the positive real axis,
 from 0 to $t/u$, and from $u/t$ to $ + \infty$.
 
 Define also functions $\lambda(z), g(z)$ by 
 \be \label{deflambda2}
 \lambda (z) +1/\lambda(z) \eq2c' c^{*} -s' s^{*} (z+1/z) \comma \ee
 which is eqn. (\ref{deflambda}), and 
\be \label{defmu2}  g(z) \eq  \frac{(1+z)[1-uz+(u-z) \rho(z) ]}
 {(1-z)[1-uz- (u-z) \rho(z)]}  \ee
 choosing $\lambda (z) > 1$ when $|z|=1$. They
are analytic and non-zero in the cut plane of Figure  \ref{posnzs}
($g(z)$  does {\em not} have zeros or poles at  $z = \pm 1$)
and   have the following symmetry properties:
 \bd \rho(1/z) = 1/\rho(z)  \sep  \lambda (1/z) = \lambda (z) \sep 
  g(1/z) = g(z)  \comma \ed
and
 \be \Delta \eq \prod_{j=1}^N \lambda (z_j) \sep 
G \eq \prod_{j=1}^N g(z_j) \period \ee

 Set
 \be \label{defhz} h(z) \eq \log g(z)  + (M-1) \log \lambda (z)  \ee
 and consider the  integral difference
  \be {\cal I}  \eq \frac{1}{2 \pi i} \left\{ \oint_{C_1} \frac{  P'(z)\, h(z) }{P(z)} 
\, dz -  \oint_{C_2} \frac{
P'(z) \, h(z) }{P(z)} \, dz \right\} \ee
where $P(z)$ is the polynomial (\ref{pol}), $C_1, C_2$ are the two circles in Figure 
\ref{posnzs}, shown as dashed
(red) lines, $z_n$ is inside $C_2$, and $z_{n+1}$ is outside $C_1$.  

The function $h(z)$ is analytic in the cut $z$-plane, which includes  $C_1, C_2$ 
and the region in between. $P(z)$ is the polynomial (\ref{pol}), so the only singularities
between $C_1$ and $C_2$  are the poles at 
$z = 1, -1,z_1,\ldots, z_{N-1}, z_{N+2}, \ldots, z_{2N}$. 

The corresponding values
of $\lambda(z)$ are $\lambda(1), \lambda(-1), \lambda_1, \ldots , \lambda_{N-1},
\lambda_{N-1}, \ldots , \lambda_1$, and similarly for $g(z)$. (Note that for $j >N,
\lambda(z_j) = 1/\lambda_{j} = \lambda_{2N+1-j}$).

It follows that $ {\cal I}  $ is the sum of the residues at these $2N$ poles, and that
\ba \label{valI}
{\cal{I}} & = &  h(1) + h(-1)  + 2 h(z_1)  + \cdots  + 2  h(z_{N-1}) \nonumber \\
 & = &  h(1) + h(-1) -2 h(z_N)  + 2 \log ( G \Delta^{M-1} )  \ea
so
\be \label{dtQ1}
\log \det Q \eq \log E  + \half [ {\cal I} +2 h(z_N) -h(1)-h(-1)] \ee

{From}  (\ref{zN}), $z_N = t/u$ to within relatively exponentially small terms in $N$,
so 
\bd \lambda (1) = \frac{(1-t)(1+u)}{(1+t)(1-u)} \sep \; \;  \lambda (-1)= 
\frac{(1+t)(1+u)}{(1-t)(1-u)}   \sep \; \;  \lambda (z_N)  = 1  \comma  \ed
\bd  
g(z_N )\,=   \frac{u+t}{u-t} \comma  \ed
\bd g(1)  \eq  \frac{u(1+u) (1-t)^2}{(1-u)(u-t)(1 - tu)}  \sep
g(-1)  \eq   \frac{(1+u)(u+t)(1+tu)}{u (1-u) \, (1-t)^2}    \ed
 taking limits $z \rightarrow \pm 1$ in (\ref{defmu2}) as necessary. It follows that
 \be \label{hhh}
 2 h(z_N) -h(1)-h(-1) \eq 2M \log \left[ \frac{1-u}{1+u} \right] + 
 \log \left[ \frac{(u+t)(1-tu)}{(u-t)(1+tu)} \right]   \ee


On $C_1$, $|z| >1$, so for $N$ large, from (\ref{pol}), neglecting only exponentially small 
terms, we can replace $P(z)$ in the integral over $C_1$ by $z^{2N} (z-ut)(z-u/t)$.

On $C_2$, $|z|<1$, so we can replace $P(z)$ in the integral over $C_2$ by $(1-tuz)(1-uz/t)$.

We make these substitutions. We can then move $C_1, C_2$ to each become the unit 
circle, giving
\be \label{integI}  {\cal I} \eq \frac{1}{2 \pi i} \; \oint h(z) \, 
\left\{ \frac{2N}{z} +\frac{1}{z-ut} - \frac{t}{u-tz } + \frac{tu}{1-tuz}  -  \frac{u}{uz-t} 
\right\}  d \, z  \ee
the integration being round the unit circle. 
	Setting $z = \e^{i \theta}$, this becomes 
the real integral
\be {\cal I} \eq \frac{1}{2 \pi} \; \int_0^{2 \pi} h(e^{i \theta}) \, 
\left\{ 2N +\sigma (\theta) \right\}  d \theta \period \ee
where 
\be \sigma (\theta) \eq \frac{1-t^2 u^2}{1+t^2 u^2 -2 t u \cos \theta } 
-\frac{u^2-t^2}{ t^2 +u^2 - 2 t u \cos \theta } \period \ee

{From} (\ref{defhz}), $h(z) $ is linear in $M$, so $\cal I$, like the RHS of (\ref{fullZ}), is
 the sum of four terms, proportional to $MN$, $M$, $N$, and 1, respectively. So are the 
 other terms that enter  $\log Z$ via (\ref{ZdetQ}), (\ref{Etot}), (\ref{dtQ1}). Define
   \be \label{defnu}
  \mu (z) \eq g(z) / \lambda (z) \comma \ee
  so that  (\ref{defhz}) becomes
  \be h(z) \eq M \log \lambda (z) + \log \mu (z) \period \ee
  
Following Onsager{\color{blue}\cite[eqn. 2.1a]{Onsager1944}}, define
  \be \label{defk}
  \kay =  ( \sinh 2H \sinh 2H' )^{-1} =  s^{*}/s' = \frac {t(1-u^2)}{u (1-t^2) } \ee
  and note that \be 1+\kay = \frac{(1+t/u)(1-ut)}{u (1-t^2) } \sep 1- \kay =
   \frac{(1-t/u)(1+ut)}{u (1-t^2) } \period \ee

Putting together (\ref{ZdetQ}),  (\ref{relnstu}), (\ref{detqEG}), (\ref{dtQ1}), (\ref{hhh}) and (\ref{integI} ),
and using the above formulae,  we obtain
\bd
\log Z \eq\frac{MN}{2}  \log 2 \sinh 2H -M H' - N (H+H^{*})  + \log 2  \; + 
\ed 
\be  \label{main} \frac{1}{8} \log \left[  \frac{ (1+\kay)^5}{(1-\kay)^3 } \right]  +
\frac{1}{8 \pi} \; \int_0^{2 \pi} \left[ M \log \lambda (e^{i \theta }) + \log \mu (e^{i \theta} ) \right]  \, 
\left[ 2N +\sigma (\theta) \right]  d \theta \period \ee
This is indeed of the expected form (\ref{fullZ}).

Separating (\ref{main}) into its four  constituent terms,  we obtain
 \be \label{bulk}
 - \beta f_b \eq \half \log ( 2 \sinh 2H) + \frac{1}{4 \pi} \int_0^{2 \pi}\log \lambda(e^{i \theta}) 
 d \, \theta \ee
 
\be - \beta f_s \eq - H' \, + \,  \frac{1}{8 \pi} \int_0^{2 \pi} 
 \sigma (\theta) \log \lambda(e^{i \theta}) d \, \theta \ee
 
\be - \beta f_s' \eq -H + H^{*} + \frac{1}{4 \pi} \int_0^{2 \pi} 
\log \mu(e^{i \theta}) \, d \, \theta \ee

\be \label{cnrfree}  - \beta f_c  \eq  \log 2 + \frac{1}{8} \,  \log \left[ \frac{(1+ \kay)^5}{(1-\kay)^3}
\right] +  
 \frac{1}{8 \pi} \int_0^{2 \pi} 
\sigma(\theta) \,  \log \mu(e^{i \theta})\, d \, \theta \period \ee

{From}  (\ref{deflambda2}), 
\be \log \lambda (e^{i \theta} ) \eq  \cosh^{-1} (c'c^{*} -s' s^{*} \cos \theta ) \ee
so (\ref{bulk}) is Onsager's famous result for the bulk free energy of the Ising 
model.{\color{blue}\cite[eqn. 106]{Onsager1944}} 


 \section{The free energies in terms of elliptic functions }
 \label{ellipfns}
  \setcounter{equation}{0}

Define $x,y$ (not to be confused with the $x,y$ of section~\ref{calcdetQ}) by 
$ \lambda (z) = (x y)^{-1} ,  z= x/y$. Then  (\ref{deflambda2}) becomes
\be x^2y^2 +1 +s's^{*} (x^2+y^2) -2 c'c^{*} xy=0 \period \ee

This is eqn. (15.10.3) of {\color{blue}\cite{book}}. I show there that such an equation can 
be parametrised using elliptic functions  of modulus $\kay$, where
\be \kay + 1/\kay =  \frac{c'^2 {c^{*}}^2 -1 -  s'^2 {s^{*}}^2}{s' s^{*} }  = 
\frac{s'^2 +{s^{*}}^2}{s' s^{*} } \period  \ee
We choose, using (\ref{exprscs}), 
\be \label{ss}
\kay =  \frac{s^{*}}{s' } = \frac{t (1-u^2)}{u (1-t^2)}  \ee
and as usual set $\kay' = (1-\kay^2)^{1/2}$.

 Let $\sn, \cn , \dn$ be the usual meromorphic elliptic Jacobi functions, and $K, K'$ the 
 complete elliptic integrals,
replacing $\eta$ in  {\color{blue}\cite{book}} by $v$, 
\be \label{cc}
c' c^{*} = - \frac{\cn v \dn v}{\kay \sn \! ^{2 \, }   v } \period \ee
{From} eqn. (15.10.12) of {\color{blue}\cite{book}}, we can introduce a parameter
$v$ such that 
$x =  - \kay^{1/2} \sn ( r-v/2 ), \, y = - \kay^{1/2} \sn (r +v/2) $, and hence
\be \label{zlam}
z = \frac{ \sn (r-v/2)}{\sn (r+v/2)} \sep 
\lambda (z) \eq \frac{1}{\kay \sn (r+v/2) \sn (r-v/2) } \period \ee

In the ordered ferromagnetic regime (\ref{tu}),  
\be 0   <  \kay < 1 \; \; {\mathrm{and}} \; \;   v, iK'-v  =  \; \mathrm{positive
 \; pure \; imaginary}  \period \ee
We  take $r$ to be real:
\be   -K  < r  \leq K  \period  \ee
Then  $\lambda$ is real and  $\lambda >1 $.
As $r$ goes from $-K$ to $K$, $z$ moves anti-clockwise around the unit circle
from 1 through $i, 0, -i$  back to 1.  

We can choose, consistently with  (\ref{ss}), (\ref{cc}),
\be  \label{cscs}
c' = \frac{i \dn v}{\kay \sn v} \sep s' = \frac{i}{\kay \, \sn v} \sep
c^{*} = \frac{i \cn v}{ \sn v} \sep s^{*} = \frac{i}{ \sn v} \comma \ee
where $\sn, \cn , \dn$ are the usual elliptic Jacobi functions. 
They satisfy the relations
\be  \label{cdrelns}
\cn \! ^2  (u) = 1-\sn \! ^2(u) \sep \dn \! ^2(u) = 1-k^2 \sn \! ^2(u) \ee
\bd \frac{d}{du} \, \sn (u) = \cn(u) \dn(u) \sep \frac{d}{du} \, \cn (u) =  -\sn(u) \dn(u)  \comma \ed
\be \label{erelns}
\frac{d}{du} \, \dn (u) =  - \kay^2  \sn(u) \cn(u) \period \ee
Setting
\be
\label{defvb} \vb= i K' -v \comma \ee
it follows that
\be \label{utclc}
u \eq -i \, \frac{\sn (\vb/2) \dn (\vb /2)}{\cn ( \vb /2) }
\sep t \eq -i \, \kay  \, \frac{\sn (\vb/2) \cn (\vb /2)}{\dn ( \vb /2) } \ee
and 
\be \label{1mt1mu}
\frac{1-t}{1+t}  \eq -i \, \frac{\sn (v/2) \dn (v /2)}{\cn ( v/2) }
\sep  \frac {1-u}{1+u} \eq -i \, \kay  \, \frac{\sn (v/2) \cn (v /2)}{\dn ( v /2) }  \ee
(interchanging  $H$ with $H'$ takes $v$ to $\vb$ and $u,t$ to $(1-t)/(1+t), 
(1-u)/(1+u)$).

One can establish many identities involving $z, t, u$ by using Liouville's theorem, as 
described in 
Chapter 15 of  {\color{blue}\cite{book}}, and writing the functions $\sn,  \cn, \dn$
as ratios of the entire theta function $H, H_1, \Theta,\Theta_1$. For instance, 
$z$ is a doubly periodic function of $r$, 
of periods $2K, 2i K'$. Because $\sn(iK'-r ) =-1/\kay \sn (r)$, it follows that 
$z$ is an even function of $r-iK'/2$, and that when $r = iK'/2$, then 
$z = - \kay \sn^2 \, ( \vb/2) = tu$. Hence $r = iK'/2$ is a double zero of the 
expression $z-tu$, and a double pole of 
\be \label{rho2}
\rho(z)^2 = \frac{(1-tuz)(uz-t)}{(z-tu)(u-tz)}  \ee
Proceeding similarly, the poles and zeros of 
\be {\cal J} \eq  - \left[ \frac{ \cn (r-iK'/2) }{\sn (r-iK'/2) \dn (r-iK'/2) }\right]^2 \ee
are also all poles and zeros of $\rho(z)^2$. There are two such double zeros, and 
two double poles. Because each factor (e.g $z-tu$) in (\ref{rho2}) 
has two zeros in a period rectangle, these are all the poles and zeros of
$\rho (z)^2$. The ratio $\rho(z)^2/{\cal J}$ is therefore entire and doubly periodic, so 
it is bounded at infinity and from Liouville's theorem must be a constant.
It is one when $r=v/2$ and $z=0$, so is one for all
$r$. Taking the square root so that $\rho (z)$ is positive real when $r$ is real, 
we have proved that
\be \label{defrho1}
\rho (z) \eq - i \, \frac{ \cn (r-iK'/2) }{\sn (r-iK'/2) \dn (r-iK'/2) } \ee
so is a  single-valued meromorphic function of $r$.

This means that the polynomial $P(z)$ in (\ref{pol}) contains the two factors
$z^n - \rho(z)$  and  $z^n + \rho(z)$, each of which is a singe-valued meromorphic
function of $r$. We used a similar factorization property in (\ref{P1P2}) of Appendix C
to calculate $E$, but  it was only in the limit of $N$ large that we could identify the two 
factors with the RH sides  of (\ref{P2J}), (\ref{P1J}). It's possible that that argument 
could be made true for finite $N$ if we had used the elliptic function parametrization.


The expression in braces on the RHS of (\ref{integI}) is
\bd \frac{d}{d \, z} \, \log [z^{2N} /\rho (z) ^2 ] \ed
so if $\theta$ is defined by $z = e^{i \theta}$, then
\be \sigma (\theta) \eq -2 z  \frac{d }{d \, z} \, \log \rho (z) \eq -2 
 \left( \frac{d }{d \, r } \log \rho \right)  
\left(  \frac{d }{d \, r } \log z   \right)^{-1}  \period \ee
Using (\ref{erelns}) and relations (8.151.2), (8.156) of   {\color{blue}\cite{GR}},
we can establish that
\be  \frac{d }{d \, r } \log \rho \eq 2 i \kay \cn (2r) \ee
and 
\be   \frac{d }{d \, r } \log z  \eq - \frac{\sn (v) [1- \kay^2 \sn\! ^2 (r+v/2) \sn \! ^2 (r-v/2) ]}
{ \sn (r+v/2) \sn(r-v/2) } \ee

Also,
\be \frac{1-uz+(u-z) \rho(z)}{1-uz- (u-z) \rho(z)} \eq \frac{i \, \lambda (z) \,  
\cn (r)}{ \sn (r) \dn (r) } \ee
\be  \frac{1+z}{1-z} \eq  \frac{\cn (v/2) \dn (v/2) \sn (r)}{\sn (v/2) \cn (r) \dn (r) } \comma \ee
so from (\ref{defmu2}  ), (\ref{defnu}),
\be \mu (z) \eq \frac{ i\, \cn (v/2) \dn (v/2) }{\sn (v/2) \dn \! ^2 (r) } \period \ee
 
 Regard $\lambda, z, \rho$ as functions of the elliptic argument $r$ and define
 \bd A_1 (r)  = \log \lambda \sep A_2 (r) = \log \kay ' -2 \log \dn  (r) \comma \ed
 \be \label{A1B2}
 B_1(r) = -i \frac{d}{dr} \, \log z \sep B_2(r) = i \frac{d}{dr} \, \log \rho \period \ee 
 
 Set \be \xi \eq \log (1-t^2)  - \half \log \kay ' +  \half \log \frac{i \cn (v/2) \dn (v/2) }
 {\sn (v/2)}  \comma \ee
  then (\ref{bulk}) - (\ref{cnrfree}) become
   \setlength{\jot}{5mm}
\ba \label{fe3}
 - \beta f_b  & = &  \half \log ( 2 \sinh 2H) + \frac{1}{4 \pi} \int_{-K}^{K} A_1 (r) 
 B_1 (r) dr \comma  \nonumber \\
 - \beta f_s  & = &   - H' \, + \,  \frac{1}{4 \pi} \int_{-K} ^{K}
A_1 (r) B_2 (r)  dr \comma \\
 - \beta f_s'  & = &  -H + \log \left[ \frac{\dn(v/2) }{\sqrt{ \kay'}} \right] +
 \frac{1}{4 \pi} \int_{-K}^{K} 
A_2 (r) B_1 (r) \, dr \comma \nonumber \\
 - \beta f_c   & = &   \log 2 + \frac{1}{8} \,  \log \frac{(1 +\kay)^5}
{(1-\kay ) ^3 }+  \frac{1}{4 \pi} \int_{-K}^{K}   A_2(r) B_2(r) dr  \period \nonumber \ea



 \subsection{The integrals as elliptic-type  sums}
 \setlength{\jot}{1mm}

 To evaluate the integrals in (\ref{bulk}) - (\ref{cnrfree}), we expand 
 $\log z, \log \rho, \log \lambda, \log \mu, $ as sums, using the product expansions
 (15.1.5), (15.1.6) of  {\color{blue}\cite{book}} and taking logarithms.
 Setting
  \be \label{defqw}
  q = e^{-\pi K'/K} \sep w = e^{i \pi v/2K}  \ee
we get
\ba \label{fourAB}
A_1(r)  & = &  \frac{ \pi (K' + i v)}{2K} +2 \, \sum_{m=1}^{\infty}
  \frac{(w^m-q^m w^{-m}) \cos ( \pi  m r /K ) }{m(1+q^m)} \comma \nonumber \\
A_2(r)  & = & - 8  \! \sum_{ m\; \mathrm{odd} }^{\infty}
  \frac{q^{m}  \cos ( \pi  m r /K ) }{m(1-q^{2m})} \comma \\
B_1(r)  & = &  \frac{\pi}{K} +\frac{2 \pi}{K} \sum_{m=1}^{\infty}
  \frac{(w^m+q^mw^{-m}) \cos ( \pi  m r /K ) }{1+q^m} \comma \nonumber \\
  B_2 (r)  & = &     \frac{4 \pi}{K} \!    \sum_{ m\; \mathrm{odd} }^{\infty}
  \frac{q^{m/2}  \cos ( \pi  m r /K ) }{1+q^m} \comma  \nonumber \ea
  where the subscript ``m odd" in the sums in the second and fourth equations means that
  the sums are over all odd  positive integers 1, 3, 5, etc. Similarly throughout this paper.
  All series in this and the next section are convergent in an annulus not smaller than
  $q^{1/2}< |w| < 1$.

Substituting these Fourier series into (\ref{fe3}), we obtain
\ba \label{res1}
 - \beta f_b  & = &  \half \log ( 2 \sinh 2H) +\frac{\pi (K'+iv)}{4 K} +\sum_{m=1}^{\infty}
\frac{ w^{2m}- q^{2m} w^{-2m}}{m (1+q^m)^2 } \nonumber \\
 - \beta f_s  & = &   - H' \, + 2 \!  \sum_{m\; \mathrm{odd} }^{\infty}
 \frac{ q^{m/2} (w^m-q^m w^{-m} )}{m (1+q^m)^2} \\
  - \beta f_s'  & = & - H +\log \frac{\dn (v/2)}{\sqrt{\kay'}} 
  - 4 \!  \sum_{ m\; \mathrm{odd} }^{\infty}  
\frac{q^m(w^m+ q^m w^{-m} )}{m (1+q^m)^2 (1-q^m) }  \nonumber \\
   - \beta f_c   & = &   \log 2 + \frac{1}{8} \,  \log \frac{(1 +\kay)^5}
{(1-\kay ) ^3 } - 8\!   \sum_{m\; \mathrm{odd} }^{\infty}
\frac{q^{3m/2}}{m(1+q^m)^2(1-q^m)} \period \nonumber \ea


Using (15.1.5) and (15.1.6) of  {\color{blue}\cite{book}},
\be \log \frac{\dn (v/2) }{\sqrt{\kay'}} \eq  2 \! 
 \sum_{ m\; \mathrm{odd} }^{\infty}
\frac{q^{m}(w^m+w^{-m})}{m(1-q^{2m})} \period  \ee
{From} (\ref {defut})
\be H^{*}  = -\half \log \tanh H = -\half \log \left( \frac{1-t}{1+t} \right)  \ee
so by using   (\ref{1mt1mu}),
\ba \label{HHH}
 \label{serHstar}  H^{*} & = &  \! \! \! \! 
 \sum_{ m\; \mathrm{odd} }^{\infty}  \frac{w^m-q^m w^{-m}}{m(1+q^m)} \comma
 \nonumber \\
H  & = &   - \frac{i \pi v}{4 K}  +\half \sum_{m=1}^{\infty} 
\frac{ q^m(w^{-2m} -w^{2m})}{m (1+q^{2m}) } \comma \\
 H '  & = &   \frac{ \pi (K'+i v)}{4 K}  +\half \sum_{m=1}^{\infty} 
\frac{w^{2m} -q^{2m} w^{- 2m})}{m (1+q^{2m}) }  \nonumber \ea
and
\be \label{shH}
 \log 2 \sinh (2H) \eq \half \log \left( \frac{4}{\kay} \right) - \frac {\pi (K'+ 2iv)}{4 K}  - 
 \sum_{m=1}^{\infty} \frac{w^{2m}-q^m w^{-2m}}{m(1+q^m)} \ee
 

Also, from (15.1.4a) and  (15.6.5) of {\color{blue}\cite{book}},
 \be \label{kapq}
  \kay' = (1-\kay^2)^{1/2} \eq \prod_{m=1}^{\infty} \left(  \frac{1-q^{2m-1}}{1+q^{2m-1}}
   \right)^4 \ee
\be \label{kapq2}
   \frac{1-\kay}{1+\kay} = \hat{\kay}' = 
\prod_{m=1}^{\infty} \left(  \frac{1-q^{m-1/2}}{1+q^{m-1/2}} \right)^4 \comma  \ee
where if we regard $\kay'$ as a function $\kay'(q)$ of $q$,
then $\hat{\kay}'  = \kay'(q^{1/2})$.
Using these formulae,  we can simplify  (\ref{res1}) to
\ba \label{res2} 
 - \beta f_b  & = &  \half \log ( 2 \sinh 2H) +\frac{\pi (K'+iv)}{4 K} +\sum_{m=1}^{\infty}
\frac{ w^{2m}- q^{2m} w^{-2m}}{m (1+q^m)^2 } \nonumber \\
 - \beta f_s  & = &   - H' \, + 2 \!   \sum_{m\; \mathrm{odd} }^{\infty}
 \frac{ q^{m/2} (w^m-q^m w^{-m} )}{m (1+q^m)^2} \\
  - \beta f_s'  & = & -H - H^{*}  +  \!  \sum_{  m\; \mathrm{odd} }^{\infty}
  \frac{(1-q^m)(w^m+q^m w^{-m} )}{m (1+q^m)^2 } \nonumber \\
   - \beta f_c   & = &       \log 2 + \frac{1}{8} \,  \log \frac{(1 +\kay)^3}
{1-\kay  } +2  \!   \sum_{m\; \mathrm{odd} }^{\infty}
\frac{q^{m/2}(1-q^m)}{m(1+q^m)^2} \period   \nonumber \ea

\subsection{Final results}
Finally, we can simplify the formulae to
\bd -\beta f_b \eq H + H' + \sum_{m=1}^{\infty} 
\frac{ q^m (1-q^m)(w^m-q^mw^{-m}) (w^{-m}-w^m)}{ m (1+q^m)^2 (1+q^{2m})}  \ed

\bd - \beta f_s = - H'  + 2  \!   \sum_{m\; \mathrm{odd} }^{\infty}
 \frac{ q^{m/2} (w^m-q^m w^{-m} )}{m (1+q^m)^2} \ed

\be \label{final}
 - \beta f'_s \eq     -H - 2  \!  \sum_{m\; \mathrm{odd} }^{\infty}
\frac{q^m(w^{m}-  w^{-m}) }{m(1+q^m)^2}    \ee

\bd  - \beta f_c \eq      \log 2 + \frac{1}{4} \,  \log \kay' +4   \!  \sum_{ m\; \mathrm{odd} }^{\infty}
\frac{q^{m/2}(1+q^{2m})}{m(1+q^m)^2(1-q^m)}  \period   \nonumber \ed
This is quite a natural way of writing the results, the preliminary terms
linear in $H, H'$ being the zero-temperature terms in a series expansion, starting
from the state(s) with all spins equal. 

For   the surface and corner free energies, this is also mathematically a natural way to write
the results, as the sums are anti-symmetric either in negating $ q^{1/2}$ while keeping
$w$ fixed, or in negating $w$ while keeping $q$ fixed; $e^{4H}$, $e^{4H'}$ and 
$\kay'$ are unchanged by such negations.

This implies some quite mysterious properties of the surface and corner free energies.
Remembering that the spontaneous magnetization of the Ising model 
is {\color{blue}\cite{Yang1952, Onsager1971, RJB2012}}
\bd {\cal M}_0 \eq \kay'^{1/4} \eq (1-\kay^2)^{1/8} \comma \ed
then considered as functions of $p = q^{1/2}$ and $w$ they satisfy
\bd 
f_s(-p,w) = f_s(p,-w) \ed
\bd  -\beta f_s(p,w)-\beta f_s(-p,w) \eq \frac{i \pi}{2} -2H'  \ed
\bd  -\beta f'_s(p,w)-\beta f'_s(-p,w) \eq \frac{i \pi}{2}   -2H \ed
\bd \exp [-\beta f_c(p)-\beta f_c(-p)]  \eq  4 \, {\cal M}_0^2  \period \ed


\subsection{Product forms}


Taking exponentials, the right-hand sides of (\ref{final}) become products. In particular,
setting
\bd p = q^{1/2} \sep s_n = \left( \frac{(1-q^{n} w^2)(1-q^{n+1} w^{-2})}{(1-q^{n})(1-q^{n+1})} 
\right)^{\! n}  \comma \ed
we find that 
\bd \negspce \! \! \! \!  \! \! \! \!  \! \! \! \!   \! \! \! \!  \! \! \! \!  e^{-\beta f_b}  =   e^{H+H'} \, 
 \prod_{n=1}^{\infty} \, 
 \frac{s_{2n-1}\,  (1-q^{4n-2})(1-q^{4n-1}w^2)(1-q^{4n} w^{-2})}{s_{2n} \, (1-q^{4n})
 (1-q^{4n-2}w^2)(1-q^{4n-1} w^{-2})}  \ed

\bd e^{-\beta f_s} \! =  \! e^{-H'}  \prod_{n=1}^{\infty} \! 
\left( \frac{(1\! + \! p^{4n-3} w)(1\! - \! p^{4n-1} w^{-1})}
{(1\! -\! p^{4n-3} w)(1\! + \!p^{4n-1} w^{-1})}   \right)^{\! \! 2n-1} \! \! 
 \left( \frac{(1\! - \! p^{4n-1} w)(1\! + \! p^{4n\! + \!1} w^{-1})}
 {(1\! + \! p^{4n-1} w)(1\! - \! p^{4n+1} w^{-1})}   \right)^{\! \! 2n} \ed
  
\bd 
e^{-\beta f'_s} \eq e^{-H} \, \prod_{n=1}^{\infty} 
\left( \frac{(1\! + \! q^{2n\! - \! 1} w^{-1})(1\! - \! q^{2n-1} w)}
{(1\! - \! q^{2n-1} w^{-1})(1 \! + \!  q^{2n-1}w)} \right)^{\! 2n-1} \!
 \left( \frac{(1\! - \! q^{2n} w^{-1})(1\! + \! q^{2n}w)}
{(1\! + \! q^{2n} w^{-1})(1 \! - \!  q^{2n}w)} \right)^{\! 2n} \ed

\be \label{prdfpsa} \!   e^{-\beta f_c} \eq 2 \, \frac{(1+\kay)^{3/8}}{(1-\kay)^{1/8}} \, 
\prod_{n=1}^{\infty} \left( \frac{1+p^{4n-3}}{1-p^{4n-3}} \right)^{4n-3}  
\!  \!  \left( \frac{1-p^{4n-1}}{1+p^{4n-1}} \right)^{4n-1} \period \ee

In the isotropic case, when
\be \label{iso}
v= i K'/2 \sep w = q^{1/4} \comma \ee
then by using (\ref{kapq}) and  (\ref{kapq2}) we can verify that
these equations agree with  Vernier and Jacobsen's
conjectures {\color{blue}\cite[eqn. 49] {VJ2012}},
except that their $q$ is our $p$ and in the last equation we have included the factor 
of 2 that comes from  the fact that for every contribution in (\ref{partfn})  to $Z$ 
from a particular configuration $\sigma$ of the spins on the lattice, there is 
another equal contribution from the spins $-\sigma$. 


\section{The inversion and rotation relations}
\label{sec-inversion}
\setcounter{equation}{0}

We emphasize that our results (\ref{final}) have been obtained by arguments that are 
rigorous, or at least could be made so. Here we present some plausible, but not rigorous,  
arguments that could have been used to obtain $f_b, f_s, f_s'$ much more easily.They 
also tell us that $f_c$ does not depend on the anisotropy parameter $v$ (or $w$).


Set \be T = V_1^{1/2}  \, V_2 \, V_1^{1/2}  \comma \ee
taking $V_1^{1/2} $ to be the positive real square root of $V_1$. Then the partition 
function is 
\be Z =  \xi^T  V_2 V_1 V_2 \cdots V_1 V_2 \, \xi \eq \xi^T  V_1^{-1/2}  \,
 T^M \,  V_1^{-1/2}  \, \xi \comma \ee
where $\xi $ is the vector with all entries $+1$. Since
\bd V_1 \, \xi \eq  (2 \cosh H )^N \, \xi \comma \ed
it follows that
\be Z =  (2 \cosh H )^{-N} \;   \xi^T T^M  \xi  \period \ee
In the physical regime, when $v$ is pure  imaginary, between 0 and $i K'$,
the matrix $T$ is real and symmetric. Hence when $M$ is large,
neglecting only terms that are relatively exponentially small in $M$,
\be Z = (2 \cosh H )^{-N} \;    \Lambda^M \langle 0 | \xi \rangle^2 \comma \ee
where $  \Lambda$ is the maximum eigenvalue of $T$ and $ | 0 \rangle $ is the
corresponding eigenvector. $M$ only enters this equation explicitly, so from (\ref{fullZ}),
 it follows that
 \be \label{breakup}
 e^{-N \beta f_b - \beta f_s } \eq \Lambda \sep e^{-N \beta f_s' -\beta  f_c }  \eq
  (2 \cosh H )^{-N} \;   \langle 0 | \xi \rangle^2 \period \ee

It has long been known that the bulk free energy of the solvable models can usually
  be obtained quite simply by the ``inversion relation method" ({\color{blue}\cite{Strog}}, 
  sections 13.6, 14.3, 14.4 of    {\color{blue}\cite{book}}. From (\ref{exprscs}), (\ref{cscs}),
  if we regard $H, H', V_1, V_2, T$  as functions of $v$, then 
  \bd H(2iK'-v) = H(v)  + i \pi /2 \sep H' (2iK'-v) = - H'(v) \ed
\be V_1(v) \, V_1(2iK' - v) = (2 \, i \sinh 2H ) ^N \mathbf{1} \sep
V_2(v) \, V_2(2iK' - v) = \mathbf{1} \comma \ee
and hence
\be  \label{invmat}
T(v) \, T(2iK' - v) = (2 \, i \sinh 2H ) ^N \mathbf{1}  \period \ee


At $v = iK$, $H =  \infty$ and $H'=0$, so both $V_1, V_2, T $ are all proportional to the 
identity matrix. Hence as $v$ moves through the point  $ iK$, the eigenvalues all become 
equal and cross over one another, the largest becoming the smallest. If $v$ is below
the inversion point $i K'$, then $\Lambda (v)$ is the largest eigenvalue and 
$\Lambda (2iK' - v)$ is the smallest. It is  reasonable to suppose that
$\Lambda (2iK' - v)$ is the analytic continuation of $\Lambda(v)$, and from
(\ref{invmat}) that
\be \Lambda(v) \, \Lambda (2iK' - v) = (2 \, i \sinh 2H ) ^N  \period \ee
{From} (\ref{invmat}), $T(2iK'-v)$ commutes with $T(v)$, so has the same
eigenvectors and  $ | 0 \rangle $ is unchanged.

This all fits with our matrix representatives calculation of section \ref{calcdetQ}. 
If $v \rightarrow 2iK'-v$, 
then $\widehat{V}_1, \widehat{V}_2$ are inverted and $\vb \rightarrow -\vb$. From
(\ref{utclc}) $u,t \rightarrow  -u,-t$ and from (\ref{pol}) we can leave $z_1, \ldots ,
z_N$ unchanged. If we also leave $\alpha_1, \ldots, \alpha_N$ unchanged, then
$\lambda_1, \ldots , \lambda_N$ are inverted and we can verify
that for each eigenvector ${\mathbf x} $ is interchanged with  ${\mathbf y} $, which 
leaves (\ref{eig}) unchanged. This is equivalent to leaving the eigenvectors of
 $\widehat{V}_1^{1/2} \, \widehat{V}_2 \, \widehat{V}_1^{1/2}$ unchanged, but inverting
 their eigenvalues.

Considering the first of the relations  (\ref{breakup}) when $v$ has its original value and 
when it is replaced by $2iK'-v$, and taking the product, we obtain relations for
$f_b$ and $f_s$. Similarly, considering the second of the relations and taking  ratios,
we obtain  relations for $f_s'$ and $f_c$. They are
\ba \label{allfour}
-\beta f_b(v) - \beta f_b(2iK'-v) & = &   i \pi/2 + \log [ 2 \, \sinh 2H ] \nonumber \\
 -\beta f_s(v) - \beta f_s(2iK'-v) & = & 0 \nonumber \\
  -\beta f_s'(v) + \beta f_s'(2iK'-v) & = &  i \pi/2 + \log \tanh H  \\
   -\beta f_c(v) + \beta f_c(2iK'-v) &  = & 0 \period \nonumber \ea

Replacing $v$ by $iK'-v$ is equivalent to interchanging $H$ with $H'$ and hence
to rotating the lattice through $90^{\circ}$, which gives the following
four relations
\ba \label{4rotns}
 f_b (iK'-v ) = f_b(v) & , &   f_s (iK'-v ) = f_s'(v) \comma \nonumber \\
f_s' (iK'-v ) = f_s(v) & , &   f_c (iK'-v ) = f_c(v) \period  \ea

We can verify that our results (\ref{final}) do indeed satisfy these relations.

Series expansions also suggest that $-\beta f_b -H -H',  -\beta f_s+H',  -\beta f_s'+H,  
-\beta f_c$  are analytic functions of  $v$, not just in the physical regime
$ 0 < \myrm{Im} \, v < K'$, but in the extended regime $ -\epsilon  < \myrm{Im} \, v
 < K' +\epsilon$, where $\epsilon$ is positive (but less than $K'$). They also suggest that
 the four functions are periodic in $v$ of period at most $4K$, so they
 are Laurent expandable in powers of $w$.
 
 These observations almost define the four free energies, as we shall now show. 
 The last  imply that there exist expansions of the form
 \ba \label{expand}
 -\beta  f_b = H + H' +\sum_{m=-\infty}^{\infty} c_{1,m} w^m & , & 
  -\beta  f_s = -H' +\sum_{m=-\infty}^{\infty} c_{2,m}w^m \nonumber \\
   -\beta  f_s' \; \; =  \; \; -H  +\sum_{m=-\infty}^{\infty} c_{3,m} w^m & , & 
  -\beta  f_c \; \;   =   \; \;   \sum_{m=-\infty}^{\infty} c_{4,m} w^m   \ea
  which are convergent for $q^{1/2} < |w| < 1 $.
  
  Substituting these into  the equations  (\ref{4rotns}) and  (\ref{allfour}) for $f_b$, 
  using the identities
  \be \log [ 2 \, \sinh 2 H] \, - 2H = \sum_{m=1}^{\infty} \frac{ (1-q^m)
   (2 q^m -w^{2m}-q^{2m} w^{-2m}) }  {m (1+q^m) (1+q^{2m} ) }\ee
   \be \log \tanh H \eq -2 \sum_{m \; {\myrm{odd}}}^{\infty} \frac{w^m-q^m w^{-m}}
   {m (1+q^m)} \comma \ee
and equating coefficients in the Laurent expansions, we obtain
\be c_{1,m} = q^{-m/2} c_{1,-m} \ee
and, for $m \neq 0$,
 \bd c_{1,m} + q^{-m} c_{1,-m}   \eq \frac{- 2 (1-q^{m/2} )}{m (1+q^{m/2})(1+q^m)} \; \;\; \; 
 {\myrm{ if }} \; m \; {\myrm{is \; even }} \comma \ed
while $c_{1,m} + q^{-m} c_{1,-m}   \eq 0$ if $m$ is odd. The case
$m =0$ gives
 \bd 2 \, c_{1,0} = 2 \sum_{n=1}^{\infty} \frac{q^n (1-q^n)}{n (1+q^n)(1+q^{2n})} \ed
 Solving these equations, we find that $ c_{1,0} = 0$ is $m$ is odd, while if $m$ is even
 and $m \neq 0$,
 \be c_{1,m} \eq  \frac{- 2 q^{m/2} (1-q^{m/2} )}{m (1+q^{m/2})^2(1+q^m)}  \period \ee
 Substituting these results(for $m$ positive, zero  and negative) 
 back into (\ref{expand}) , we obtain the result (\ref{final}).
 
 Similarly, using  (\ref{4rotns})  and (\ref{allfour})  for $f_s, f_s'$, we get the equations
 \bd c_{3,m} = q^{-m/2} c_{2,-m} \sep c_{2,m} +q^{-m} c_{2,-m} =0 \comma \ed
 \bd  c_{3,m} -q^{-m} c_{3,-m} = \frac{\! \! \! \! \! \!   - \, 2}{m (1+q^m)} \; \; \; {\myrm{if} } 
 \; m \; \myrm{is \; odd}  \sep = \;  0\; \; \; {\myrm{if} } 
 \; m \; \myrm{is \; even} \period \ed
 Solving these gives 
 \be c_{2,m} = \frac{ 2 \, q^{m/2}}{m (1+q^m)^2} \sep
  c_{3,m} = \frac{ - 2 \, q^{m}}{m (1+q^m)^2}\; \; \; {\myrm{if} } 
 \; m \; \myrm{is \; odd}  \comma \ee
 while $c_{2,m} = c_{3,m} = 0$ if $m$ is even. This also agrees with (\ref{final}).
 
 Finally, for $f_c$,  (\ref{4rotns})  and (\ref{allfour})  give
 \be c_{4,m} \eq q^{-m/2} \, c_{4,-m} \eq  q^{-m} \, c_{4,-m}  \period \ee
These equations imply that 
 \be c_{4,m} =0  \; \; \; {\myrm{if} } 
 \; \; m \neq 0  \comma \ee
 so $f_c$ is indeed independent of $v$ and $w$, as we found. These arguments do
 {\em not} give the value of $c_{4,0}$, i.e. the constant term in the Laurent expansion
 (\ref{expand}) of $f_c$.

Our derivation in this section is {\em not} rigorous, because we have assumed the
existence of the Laurent expansions  (\ref{expand}). For the Ising model on the
square lattice rotated through $45^{\circ}$, with cylindrical boundary conditions,
the row-to-row transfer matrices {\em commute} (because of the Yang-Baxter
relations). This means that their eigenvalues, like the Boltzmann weights
$\e^{-2H}, e^{-2H'}$, are meromorphic functions of $v$, even for a finite 
number $N$ of columns. One can then establish rigorously an inversion
identity, and from that calculate $f_b$.

However, we have used closed boundary conditions and the orientation of Figure
\ref{sqlattice1}, so our transfer matrices do {\em not} commute. (The eigenvectors
depend on $z_1, \ldots, z_N$, the zeros of the polynomial $P(z)$ in
(\ref{pol}), which certainly depends on $v$.) We do not have any \textit{a priori}
reason for believing the surface and corner free energies to be meromorphic functions 
of $v$.

Indeed, for unsolved models, such as the Ising model in a magnetic field,
one can establish inversion  and rotation relations  like (\ref{allfour}) and 
(\ref{4rotns}), but the free energies
have complicated singularities at the inversion points and one does not have
useful expansions like (\ref{expand}).

Having said this, O'Brien, Pearce, 
Behrend and Batchelor{\color{blue}\cite{Pearce95, Batchelor96,Pearce97}} 
have obtained surface free energies for various solved models by using
the ``reflection Yang-Baxter relations".  These lead to commutation 
properties of transfer matrices, and Pearce has used the resulting inversion identities
to obtain the surface free energies of the self-dual Potts model. The relation 52 of his notes
{{\color{blue}\cite{Pearcenotes}} } is so like our relation (\ref{allfour}) for $f_s, f_s'$
that it must be possible to obtain a more rigorous derivation by his methods.


 \section{The corner  free energy}
 \label{cornerfreeenergy}
\setcounter{equation}{0}

The non-rigorous, but comparitively simple, arguments of the
previous section do not give us any information on the corner free
energy $f_c$, beyond telling us that it does not depend on the anisotropy
parameter $v$.
 
 However, there is one intriguing point that gives some hope that it may be
 possible to obtain it, at least to within simple additive algebraic functions of $k$.
 
The integrands in the four equations of (\ref{fe3})  are 
$A_1B_1, A_1B_2, A_2 B_1,  A_2 B_2$, respectively.
This directly leads to the fact that
\be \label{ssss}
S_1 S_4 = S_2 S_3 \comma \ee
where $S_1, S_2, S_3, S_4$
are the summands (including the external numerical factors)
in the series in the four equations (\ref{res1}).

The same is true of the summands in (\ref{res2}) and (\ref{final}). (The second form of
the equations be obtained, at least formally,  by re-defining the function $A_2(r)$, 
while the third needs a re-definition of both  $A_2(r)$ and $B_1(r)$.)
Thus if one can obtain the bulk and surface free energies, this will give 
$S_1, S_2, S_3$, so $S_4$ can then be obtained from (\ref{ssss}).
Considered as functions of $q$, these summands all have a double pole
at $q^m=-1$, and it this double pole that makes the free energies 
non-algebraic functions of the Boltzmann weights. It ensures that they are products
of factors such as $1-q^{2n} w$,  with exponents that are {\em linear}  in  $n$
 (rather than constants),  as in (\ref{prdfpsa}).



\section{Critical behaviour}
\label{sec-critical}
\setcounter{equation}{0}

\subsection{Bulk free energy}

The Ising model is critical when $\kay =1$ and $q=1$. We can  obtain the 
behaviour near criticality
by using the Poisson tranform given in (15.8) of {\color{blue}\cite{book}}:
\be \label{Poiss}
\delta  \sum_{n=-\infty}^{\infty}   f(n \delta) \eq  \, \sum_{n=-\infty}^{\infty} 
g( 2 \pi n /\delta ) \comma \ee
where
\be\label{Four}
 g(k) \eq \int_{-\infty}^{\infty} e^{ikx} \, f(x) \, dx \comma \ee
true for any function $f(x)$ that is analytic on the real axis and for which the integral
is absolutely convergent.

If we take 
\be \label{deffx}
f(x) \eq \frac{ \pi \sinh  2  \alpha x/\pi}{2 \alpha \, x \,  (\cosh x )^2} \comma \ee
then $f(x), g(k)$ are even functions, so we can write (\ref{Poiss})
as
\be \label{Poiss2}
\delta  [ f(0) +2 \sum_{n=1}^{\infty} f(n \delta) ] \eq g(0) +2  \, \sum_{n=1}^{\infty} 
g( 2 \pi n ) \comma \ee

For $k > 0$ we can  close the integration in 
(\ref{Four}) round the upper half $x$-plane and sum over the residues of the poles
at $x = i (2m-1) \pi/2$. This gives $g(k)$ as a sum over $m$, the summand
being a sum of terms that are either exponential in $k$, or proportional to the same 
exponential multiplied by $k$.

It follows that we can perform the summation on the RHS of (\ref{Poiss2}) for each pole, 
giving
\be \label{ident}  \delta +  \sum_{n=1}^{\infty} 
\frac{\pi \sinh (2 \alpha \delta n/ \pi)}{\alpha \, n  \, \cosh^2 (\delta n) }
\eq g(0) +R_1 +R_2 \comma  \ee
where  (replacing $2m-1$ by $n$)
\ba   \setlength{\jot}{5mm}
R_1 & = &   8 \sum_{n\; \myrm{odd} }^{\infty} \frac{y^{2n}}{n (1-y^{2n})} 
 \left[  \frac{ \sin ( n \alpha)}{n  \alpha} -
\cos  ( n  \alpha) \right]  \comma \nonumber \\
 R_2  & = &  \frac{8 \pi ^2}{\alpha \, \delta } \, \sum_{n \; \myrm{odd} }^{\infty}   
\frac{y^{2n} \sin (n   \alpha) }{n (1-y^{2n})^2} \;  \comma \ea
and
\bd y = e^{-\pi^2/2 \delta } \period \ed

Setting
\be\label{defda}
 \delta = \frac{\pi K'}{2 K} \sep \alpha = \frac{\pi (K'+iv)}{K'} \comma
\ee 
we see that $\alpha$ is real,  $ \pi/2  < \alpha  < \pi $, and the first equation of  (\ref{res1})
can be written
\be \label{ffb}
- \beta f_b \eq \half \, \log( 2 \sinh 2H) +\frac{ \alpha I}{2 \pi}   \comma \ee
where $I$ is the LHS of (\ref{ident}) and
 \bd y  = e^{-\pi K/K' } = {q'} \period \ed
 
 The function $k' (q')$ is the same as $k(q)$, so from (15.1.4a) of {\color{blue}\cite{book}},
 \bd k' = 4\, {q'}^{1/2} \prod_{n=1}^{\infty} \left( \frac{1+{q'} ^{2n}}{1+{q'} ^{2n-1}}\right)^4 
 \period  \ed
 
 Let $\Delta T = T_c - T$, where $T$ is the temperature and $T_c$ its value at criticality.
 Then near criticality $1-k$ vanishes and is proportional to $\Delta T $. Hence so are 
 $k'^2 =1 -k^2$ and $q'$. In fact $q'$ is an analytic function of $T$, with a simple zero 
 at $T_c$.
 
 The parameter $\alpha$ is non-zero and analytic at $T_c$, while 
 $\delta = - \pi^2/(2 \log q') $
 diverges logarithmically. Substituting the RHS of   (\ref{ident}) for $I$ in  (\ref{ffb}),
 we find that the only term that is not analytic is the last, i.e. $\alpha R_2/2 \pi$, and the 
 dominant singular contribution to $- \beta f_b$ is
 \bd  \frac{4 \pi {q'}^2 \sin \alpha }{\delta }  \eq   - \frac {8\, 
 {q' }^2 \log q'  \,  \sin  \alpha  }{\pi}   \period  \ed

 The internal energy is proportional to the first derivative of $f_b$ with respect to
 $T$, and the specific heat to the second derivative. We see that the bulk free energy 
 and internal energy are finite when $T=T_c$, but the
specific heat diverges logarithmically, as found by 
 Onsager {\color{blue}\cite{Onsager1944}},  {\color{blue}\cite[eqn (7.12.10)]{book}}.
 
 \subsection{Surface and corner  free energies}
 
 We can restrict the sum in (\ref{Poiss}) to even $n$ by replacing $\delta$ by 
 $2 \delta$ and  dividing by 2. If we 
 then subtract the result from the original equation we obtain another identity:
\be   \label{ident2} \sum_{n \; \myrm{odd} }^{\infty} \frac{ \sinh (2 \alpha \delta n/\pi)}{n  \, 
\cosh^2 (\delta n) }  \eq \alpha \left[ \half g(0) +R'_1 +R'_2 \right]/ \pi \comma  \ee
where
\ba R'_1 & = &   - 4 \sum_{n \; \myrm{odd} }^{\infty} \frac{y^n }{n (1+ y^n )} 
 \left[  \frac{ \sin ( n  \alpha)}{n \,  \alpha} -
\cos  ( n \alpha) \right]  \comma \nonumber \\
 R'_2  & = &  - \frac{2 \pi^2 }{\alpha \, \delta } \, \sum_{n\; \myrm{odd} }^{\infty}   
\frac{y^n \sin (n  \alpha) }{n (1+y^n)^2} \;  \comma \ea
and $g(x), y$ are defined as before.
In these equations, as throughout this paper, the subscript  ``${n \; \myrm{odd} }$" 
 means that the sum is over all odd {\em positive}  integer values of $n$, i.e.
  $n = 1, 3, 5,$ etc.
  
  If we now define $\alpha$, not by (\ref{defda}), but by
  \be \alpha \eq  \frac{\pi (K'+iv)}{2K'} \comma \ee
  then the last term in the second equation of (\ref{res1}) is the LHS
  of  (\ref{ident2}), so
  \be -\beta f_s \eq -H' +\alpha \left[ \half g(0) +R'_1 +R'_2 \right]/ \pi \period  \ee
  As for the bulk free energy, all the terms on the RHS of this equation are analytic functions
  of  $q'$ at $q' = 0$, except for the $\delta$ in $R'_2$, so the dominant 
  singularity in $= -\beta f_s$ is
\be \label{logsing}  \frac{q' \log q' \sin \alpha}{\pi}   \ee
and we see that the first derivative of $f_s$ (the ``surface internal energy") diverges
logarithmically. Since $f_s$, $f'_s$ differ only in replacing $v$ by $\vb = iK' -v$ and $\alpha$
by $\pi/2 - \alpha$, the same is true of $f'$.

The sum in the last equation of (\ref{res2}) is the same as that in the second, but with $w=1$,
i.e. $v=0, \alpha = \pi/2$, so that has the same logarithmic singularity as  (\ref{logsing}).
However, the second term the equation contains a contribution
\bd  - \frac{\log (1-\kay) }{8}   \ed
Since $1-\kay = 4 q'$ when $q'$ is small, the corner free energy itself diverges 
logarithmically near criticality. 

If we define a critical exponent $\widehat{\alpha}$
in the usual way   {\color{blue}\cite[eqn 1.7.10b]{book}} so that the free energy
near $T = T_c$ has a singularity proportional to $(T_c-T)^{2-\widehat{\alpha}}  $, or 
in this case (where $\widehat{\alpha}$ is an integer)
\bd  (T_c-T)^{2-\widehat{\alpha}}  \,  \log (T_c-T) \ed
then $\widehat{\alpha}$ has the values $0,1,2$ for the bulk, surface and corner 
free energy, respectively.


\section{Summary}
\setcounter{equation}{0}

Prompted by the conjectures of Vernier and Jacobsen{\color{blue}\cite{VJ2012}} for the
 isotropic case,  we have used the spinor method of 
 Kaufman{\color{blue}\cite{Kaufman1949}} to calculate 
the bulk, surface  and corner free energies of the two-dimensional anisotropic
Ising model. We do indeed find agreement with 
{\color{blue}\cite{VJ2012}} for the isotropic case. 

The bulk free energy was calculated by Onsager in 
1944{\color{blue}\cite{Onsager1944}}, and the surface free energy by McCoy and 
Wu in 1967 {\color{blue}\cite[eqn.4.24b]{McCoyWu67}\cite[p.126, eqn.4.24b]{MCWbook}} 
We have used Kaufman's method to calculate these, together with the corner free energy $f_c$.
This last  is  by far the most difficult to obtain, involving(unlike  $f_b$ and $f_s$), the
calculation of $E$, $G$ and $A_2(r)$, to which much of this paper, 
including the whole of the two appendices, is devoted.

We emphasize that our full derivation is rigorous (or at least could be made so), except in 
section \ref{sec-inversion}. The purpose of that section is to show how one can obtain 
$f_b, f_s, f_s'$ much more easily if one is prepared to make assumptions about the their
analyticity properties, and that one also can show that $f_c$ is independent of the 
anisotropy parameter  $v$ (or $w$). In that respect $f_c$ (for the rectangular lattice)
is similar to the order parameter (i.e. the  spontaneous  magnetization)  ${\cal M}_0$.
  
 
An intriguing point to which we have referred is that there is structure in the
 four equations (\ref{fe3}) for $f_b, f_s, f_s', f_c$, coming from the fact that the integrand in
 (\ref{main}) is a product of two terms, one linear in $M$, the other in $N$. This means that
 the free energy summands $S_1, S_2, S_3, S_4$  in  each of (\ref{res1}),
 (\ref{res2}) and  (\ref{final}) satisfy the relation (\ref{ssss}), i.e. $S_1 S_4 = S_2 S_3$, so 
 the summand for the corner free energy is obtainable, using this relation, from
 those for the bulk and surface free energies. If this could be justified, and the additional
 simple algebraic terms in, say, equation (\ref{final}), explained, then we could eliminate
 the lengthy calculation herein of the corner free energy.
 
 \section{Acknowledgements}
 The author is is extremely grateful to Barry McCoy for pointing that he and T. T. Wu 
 obtained the surface free energy $f_s$ in 1967. Their book was sitting on the shelf 
 behind me as I wrote most of this paper, and clearly I should have consulted it at the 
 beginning. The author also thanks Paul Pearce and Murray Batchelor for their timely 
 reminders of the work on the reflection Yang-Baxter relation and its use in handling the
 transfer matrices of lattices with closed boundary conditions, and to Paul Pearce for 
 sending him his notes on a derivation of the surface free energy of the self-dual Potts
 model. The author  is indebted to Helen Au-Yang and Jacques Perk for
 alerting him to the work on surface and corner magnetizations, and for spotting some 
 typographical errors. He is grateful to Paul Fendley for helpful comments.


 \appendix

\renewcommand{\theequation}{A\arabic{equation}}
\setcounter{equation}{0}

\section{The formula  of McCoy and Wu }
The relevant low-temperature result of McCoy and Wu is given in eqn.
(4.24b) of {\color{blue}\cite{McCoyWu67}},  and in eqn. (4.24b) of page 126 of 
 {\color{blue}\cite{MCWbook}}. Taking 
$\beta E_1, \beta E_2$ therein to be our $H, H'$, defining
\be z_1 = \tanh H \sep z_2 = \tanh H' \sep  \alpha_1 = z_1 e^{-2H'} \sep 
  \alpha_2  = e^{-2H'}/z_1 \comma \ee  
 and making some minor adjustments of 
notation, it is
\be \label{McWres}
- \beta f_s \eq - \log [2 \cosh H' \, ] -  \frac{i}{\pi} \int_{-i \infty}^{+ i \infty} 
\frac{d \omega}{1-\omega^2} \log [1 - U^{1/2} ] \comma \ee
where \be U = \frac{(\tau_1-\omega) (\tau_2-\omega)}{(\tau_1+\omega) (\tau_2+\omega)} 
\sep \tau_1 \eq \frac{1-\alpha_1}{1+\alpha_1} \sep \tau_2 \eq
 \frac{1-\alpha_2}{1+\alpha_2} \period  \ee
McCoy and Wu have a factor $1/2$ inside the second logarithm in (\ref{McWres}): we
 have shifted its contribution into the first logarithm, where it gives the factor 2. The square 
 root is to be chosen so that $U^{1/2}$ is one when $\omega = 0$, and continuous on and 
 near the imaginary axis. The symbols 
 $\omega, z_1, z_2, \alpha_1, \alpha_2, \tau_1, \tau_2, U$ used in this 
 Appendix are {\em not} to be confused with any similar symbols elsewhere in this paper.
 
 We fix the definitions of the integrands in (\ref{McWres}) and (\ref{defcJ} ) below by 
 first shifting the contour of integration  to just to the right of the imaginary axis (thereby 
 avoiding the logarithmic singularities at $\omega = 0$ and $\infty$), requiring 
 $\log (1-U^{1/2})$ and $\log (1+U^{1/2})$  to be 
 real when $\omega$ is real and  $ 0 < \omega < \tau_1$, and to be continuous in 
 the right-half $\omega$-plane, except only across the cut on the real axis from $\tau_1$ 
 to $\tau_2$. This ensures that the integrand when $\omega$ is 
 conjugated is the conjugate of the original  integrand, and the integrals (after including the 
 multiplication by $i$) are real.
 

 
 We rewrite $\log [1 - U^{1/2}]$ as
 \be \label{2terms}
 \half \log [ 1-U] + \half \log \frac{1-U^{1/2}}{1+U^{1/2}} \ee
 and note that 
 \bd 1-U = \frac{2 \, (\tau_1 +\tau_2) \, \omega }{(\tau_1+\omega) (\tau_2+\omega)}
   \period \ed
   
  Consider the contribution of the first term in (\ref{2terms}) to the RHS of
  (\ref{McWres}). The function $\log (1-U) $ is analytic in the RH $\omega$-plane. We can 
  close the contour  of integration round the RH plane and the only singularity is a simple 
  pole at $\omega = 1$, with residue
  \bd  - \frac{i }{4\pi} \log (1-\alpha_1 \alpha_2 ) =  - \frac{i}{4\pi} \log (1-e^{-4H'} ) \ed
  Noting that the integration is clockwise round the contour, we obtain
  a contribution to   (\ref{McWres}) of $ \half \log (1-e^{-4H'} ) $ and (\ref{McWres}) 
  becomes
  \be \label{McWres2}   -  \beta f_s \eq   - H'  +   \half \log \tanh H' - 
 { \cal {J}} \comma \ee
 where \be \label{defcJ}
 {\cal J} \eq 
 \frac{i}{2 \pi} \int_{-i \infty}^{+ i \infty} 
\frac{d \omega}{1-\omega^2} \log \left[ \frac{1 - U^{1/2} } {1 + U^{1/2} } \right] \comma \ee

 If we define $\chi$ so that 
 \be \omega = \tau_1 \, \frac{\chi - 1}{\chi + 1 } \comma \ee
 then as $\omega $ goes from $  - i \infty $ to $  + i \infty $ along the imaginary axis, 
 $\chi $ goes anticlockwise  once round the unit circle, from $-1$ through  $-i , 1 , i $ back  to  
 $-1$.
 
 {From} (\ref{defut}), $ z_1  = (1-t)/(1+t)$ and $z_2  = u$.
 Use the elliptic parametrization of section \ref{ellipfns}, with modulus $k$. Then it follows 
 from (\ref{1mt1mu})
 that \be \label{defalpha1}
 \alpha_1 = -k \sn^{\! 2} (v/2) \sep \alpha_2 =  k \frac{\cn^{\! 2} (v/2)}{\dn^{\! 2}(v/2)}
 \period \ee
 and we can verify that
 \be \tau_2 = \tau_1 \, \frac{1-k}{1+k} \period \ee
 
 If we define $r$ so that
 \be  \label{defchi}
 \chi^{-1}  =  k \, \sn^{\! 2} r \ee
 then
 \be \label{Drt} U^{1/2} \eq \frac {i \,  k \sn r \, \cn r}{\dn r  } \period \ee
 As $r$ moves along the horizontal line from $i K'/2$ to $2K +i K'/2$ in the complex
 plane, $\chi$ and $U^{-1/2}$ both go anti-clockwise once round the unit circle, from 
 $-1$ through   $-i, 1, i$  back to $-1$, and $\omega$ goes upwards along the imaginary 
 axis from $-i \infty$ to $+ i \infty$.

 We can verify, using (\ref{defalpha1}) and (\ref{defchi}) and methods similar to those used 
 to obtain (\ref{rho2}) herein, that
 \bd  \frac{1 - U^{1/2} } {1 + U^{1/2} }  \eq \frac{-i \cn (r \minus iK'/2)} {\sn (r \minus iK'/2)\, 
 \dn (r \minus iK'/2)} \eq \rho \comma \ed
\be \frac{1-\omega}{1+\omega} \eq k \sn (r-v/2) \sn (r +v/2) \eq \lambda^{-1}  \comma \ee
where $\rho,\lambda$ are the $\rho, \lambda $ defined as  functions of $r$ in 
 (\ref{defrho}) and (\ref{zlam}).


This almost completes the identification of our result for $f_s$ with that of McCoy and Wu.
Using the definitions in  (\ref{A1B2}):
\bd A_1 (r)  = \log \lambda  \sep B_2(r) = i \frac{d}{dr} \, \log \rho \comma \ed
we see that
\bd  \frac {d \, \omega } {1-\omega^2} \eq \half A_1'(r) \, dr \comma \ed
and we can write (\ref{defcJ}) as
 \be \label{defcJ1} {\cal J}  \eq 
  \frac{i}{4 \pi} \int_{iK'/2-i \epsilon }^{2K+iK'/2 - i \epsilon} 
A_1'(r)  \, \log \rho(r)  dr \comma \ee
where $\epsilon$ is small and positive real (this corresponds to the shift of the
$\omega$-integration to just to  the right of the imaginary axis).


The integrand is a periodic function of $r$, of period $2K$, 
so we can shift the integration down to the real axis,
 but have to allow for the simple pole in $A_1'(r)$ at $r = v/2$, where,
  using  (\ref{defvb}), (\ref{utclc}) and (\ref{defut}),
 \be \rho (r) = \rho(v/2) =  \frac{ i \cn \! (\vb/2)}{\sn \! (\vb/2) \dn \! ( \vb/2) }  \eq  
 u^{-1} \eq  \coth H'  \period \ee
Also shifting the contour back 
 a distance $K$, we get therefore
 \be  {\cal J}  \eq \half \log \tanh H' + 
  \frac{i}{4 \pi} \int_{-K}^{K} 
A_1'(r)  \, \log \rho(r)  dr \period \ee

Substituting this result into (\ref{McWres2} ) and integrating by parts (using 
the fact that both  $\log \lambda$ and $\log \rho$ are now periodic of period $2K$), we 
find that  McCoy and Wu's  result (\ref{McWres}) is equivalent to
\be  - \beta f_s \eq  - H'  + \frac{1}{4\pi} \int_{-K}^{K}
A_1(r) B_2(r) dr \comma \ee
and this is indeed the same as our result (\ref{fe3}).



\setcounter{equation}{0}

\section{The determinants of $Y$ and $C$.}
Throughout this and the following appendix, $k$ is an integer suffix,
 {\em not} the elliptic modulus introduced in (\ref{defk}).

\renewcommand{\theequation}{B\arabic{equation}}
\renewcommand{\thesubsection}{B.\arabic{subsection}}

\subsection{The determinant of $Y$}

To calculate the determinant of the $2N$ by $2N$ matrix $Y$, we first note that it
can be transformed to  a matrix composed of two $N$ by $N$ diagonal blocks:
\be\label{eqA1}
 \widehat{Y} \eq \half [ {\mathbf 1}+(-1)^N i S ] \, Y  {\cal B} \ee
where $S$ is defined by (\ref{defWS}) and $\cal B$ is the matrix that 
puts all the odd columns into positions $1, 2, \ldots , N$, followed by all the even columns.
Thus
\bd {\cal B}_{k',k}  =1  \; \; \myrm{ if } \; \; k' = m(k) \sep {\cal B}_{k',k}   = 0 \; \;
 \myrm{ else } \comma \ed

where 
\bd m(k) = 2k-1 \; \;  \myrm{ if } \; \; k \leq N  \sep  m(k)  = 2k-2N \; \;  \myrm{ if } 
\; \; k > N  \period \ed

{From} (\ref{y}), it follows that, for $j =1 , 2, \ldots, N$ and  $ 1\leq k \leq 2N$,
\be \label{Y1} \widehat{Y}_{j,k} \eq \half  \rho_k \, [z_m^{j-1} -z_m^{2N+1-j}+ 
u (-1)^N(z_m^{N+j-1}-z_m^{N+1-j})]
\comma \ee
\be \label{Y2}  \widehat{Y}_{j+N,k} \eq \half  i \rho_k'  \, [u \, z_m^{j} - u \, z_m^{2N-j}- 
(-1)^N(z_m^{N+j}-z_m^{N-j})]
\comma \ee
where $m = m(k)$, $\rho_m= 1-(-1)^N \alpha_m $, $\rho_m'= 1+(-1)^N \alpha_m$.

Using (\ref{alphak}), it follows that $\rho_k = 1-(-1)^{m(k)}$,  so is zero for $k>N$.
Similarly, $\rho_k' =0 $ if $k \leq N$. Hence $\widehat{Y}$ is a block-diagonal matrix, 
with two $N$ by $N$  blocks.

First consider the top-left block $Y_1$, with elements given by (\ref{Y1}) for 
$1 \leq j,k \leq N$ and $\rho_k=2$. For small values of $N$ we find
(using Mathematica), for arbitrary $z_1, z_2, \ldots, z_{2N}$,  that
\be \label{detY1}
\det Y_1 \eq  c_1 \, \prod_{j=1}^N (1-z_{2j-1}^2) \; \prod_{1 \leq j <k \leq N} 
(z_{2j-1}-z_{2k-1})(1-z_{2j-1} z_{2k-1}) \comma
\ee
where  \ba \label{valc1}
c_1  & = &   (u-1) \, (u^2-1)^{N/2-1}  \; \; \myrm{if} \; N \; \myrm{is \; even} 
\comma  \nonumber \\
 & = &   (u^2-1)^{(N-1)/2} \; \; \myrm{if} \; N \; \myrm{is \; odd}  \period  \ea

 We can prove that this is correct for all $N$. Consider the first column of $Y_1$.
Its elements are all polynomials in $z_1$, of degree at most $2N$, and $z_1$ occurs 
only in this column.
The determinant must therefore be a 
 polynomial of this degree  in $z_1$. If, for $k>1$, $z_1= z_{2k-1}$, then columns 
 $1$ and $k$ are identical
 and the determinant vanishes, so $z_1- z_{2k-1}$ is a factor of this polynomial.
 
 Also, inverting $z_1$ merely divides all the elements of the column 1 by $-z_1^{2N}$,
  so  $1-z_1 z_{2k-1}$ must also be a factor.
  
  If $z_1 = \pm 1$, all elements of the first column vanish, so $1-z_1^2$ must be  a 
  factor.
  
  Hence the determinant must be of the form
  \bd \tilde{c}_1 (1-z_1^2) \prod_{k=2}^N (z_1-z_{2k-1})(1-z_1 z_{2k-1}) \ed
  where $\tilde{c} _1$ is a polynomial in $z_1$. But the other terms have total degree 
  $2N$, so $\tilde{c} _1$ must be independent of $z_1$.
  
  Repeating this argument for the other columns, the determinant must be of the 
  form(\ref{detY1}),   where $c_1$ is independent of $z_1, z_3 \ldots, z_{2N-1}$.
  
  To calculate $c_1$, left-multiply $Y_1$ by a lower-triangular matrix $\cal M$ that ensures 
  the exponents of the ${Y_1}_{j,k}$ monotonically decrease as $N$ increases. 
  This does not change the determinant.
  
  If $N$ is odd,
  we take  ${\cal M}_{j,j}= 1 $, and ${\cal M}_{j,N=2-j}= -u $ provided $2j<N-2$, else 
  ${\cal M}_{j,k}=0$. Then the terms of highest power in  $z_{2k-1}$ in column $k$ are
  \be \label{terms}
   - z^{2N}, -z^{2N-1},   \ldots    , -z^{(3N+1)/2}, (u^2 -1) z^{(3N-1)/2}, 
  \ldots ,(u^2 -1) z^{N+1} \ee
  
The term in the determinant that is proportional to $z_1^{2N} z_3^{2N-1}
  \cdots z_{2N-1}^{N+1}$ is the product of the diagonal elements of ${\cal M } Y_1$ and from 
  our last result this has coefficient 
  \bd (-1)^{(N+1)/2} \, (u^2-1)^{(N-1)/2} \period \ed
  The corresponding term on the RHS of (\ref{detY1}) is $(-1)^{(N+1)/2} c_1$. They 
  must be equal, so $c_1 = (u^2-1)^{(N-1)/2}$, which is the formula given in 
  (\ref{valc1}) for $N$ odd.

  The proof for $N$ even is similar, the main difference being that the term in position
  $N/2+1$ in the sequence (\ref{terms}) is $(u-1) z^{3N/2}$.
  
{From}  (\ref{Y1}), for $j,k =1 , \ldots N$,
  \be \widehat{Y}_{2N+1-j,k+N} \eq i (-1)^N  \{ \widehat{Y}_{j,k}  \} \ee
  where $ \{ \widehat{Y}_{j,k}  \}$ is the RHS of (\ref{Y2}), but with $u, z_{2k-1}$
  replaced by $-u, z_{2k}$.
   It follows that if $Y_2$ is the lower-right block of $\widehat{Y}$, then

  \be \label{detY2}
\det Y_2 \eq  c_2 \, \prod_{j=1}^N (1-z_{2j}^2) \; \prod_{1 \leq j <k \leq N} 
(z_{2j}-z_{2k})(1-z_{2j} z_{2k}) \comma
\ee
where  \ba \label{valc2}
c_2  & = &   -(u+1) \, (u^2-1)^{N/2-1}  \; \; \myrm{if} \; N \; \myrm{is \; even} 
\comma  \nonumber \\
 & = &   - i  (u^2-1)^{(N-1)/2} \; \; \myrm{if} \; N \; \myrm{is \; odd}  \period  \ea
 
 Multiplying (\ref{detY1}) and  (\ref{detY2}) together, and noting from (\ref{eqA1}) that
$\det \widehat{Y} = (-1)^{N(N-1)/2} \, 2^{-N} \det Y$, we obtain
\be \det Y \eq (-2 i)^N (1-u^2)^{N-1} \prod_{j=1}^{2N}  (1- z_j^2)
 \prod_{1 \leq j < k  \leq 2N}^{**}
(z_{j}-z_{k})(1-z_{j} z_{k}) \comma \ee
where the $^{**}$ on the second product means that it is over all pairs $j,k$ of the 
same parity (both even or both odd).

Now we use the fact that $z_{2N+1-k} = 1/z_k$ to reduce this result to the
form
\be \label{resY}
\det Y \eq  2^N \epsilon_N \, (1-u^2)^{N-1} \prod_{j=1}^N 
\frac{(1-z_j^2)^2}{z_j^{2N}}
\, \prod_{k=1}^{j-1} (z_j-z_k)^2 (1-z_j z_k)^2
\period \ee

\subsection{The determinant of $C$}
{From} (\ref{defC}) and (\ref{defCF}), $C$ is the $N$ by $N$ matrix with elements
\be \label{elemC}
C_{jk} =  = z_k^{j-1}+(-1)^{N-k} \, z_k^{N-j} \period \ee
For small values of $N$ we find that
\be \label{detc}
\det C \eq \tilde{c}  \prod_{1 \leq j < k  \leq N}^{**}
(z_{j}-z_{k})(1-z_{j} z_{k}) \comma \ee
where again the  $^{**}$  means that the product is restricted to $j,k$ such that 
$j-k$ is even, and
\ba \label{tc}
 \tilde{c}  & = &   2^{N/2} \prod_{j=1}^{N/2} (1-z_{2j-1})(1+z_{2j})  
\; \; \myrm{if} \; \; N \; \; \myrm{is \; even} \nonumber \\
& = &   2^{(N+1)/2} \prod_{j=1}^{(N-1)/2} (z_{2j}^2-1)
\; \; \myrm{if} \; \; N \; \; \myrm{is \; odd} \period  \ea

We prove that (\ref{detc}) is correct for all $N$ in a similar way to our proof above
of (\ref{detY1}). The determinant of $C$ is a polynomial in any $z_j$, of  degree $N-1$.
If, for two $j,k$ of the same parity,  either $z_j = z_k$ or $z_j = 1/z_k$, 
then columns $j,k$ of the matrix $C$ will be proportional to one another, so the 
determinant of $C$ will vanish. Hence $\det C$ contains the product in (\ref{detc}) as a factor.  
This product is a polynomial if each $z_j$ of degree $N-2$ if $N$ is even. if $N$ is odd,
it is of degree $N-1$ in $z_k$ for $k$ odd, of degree $N-3$ in $z_k$ for $k$ even. 

If $n-k$ is odd and $z_k =1$, then $C_{jk}=0$ for all $j$, so again  $\det C$
vanishes.
Similarly if $k$ is even and $z_k= -1$.  It follows that $\tilde{c}$ must contain the products
in (\ref{tc}) as factors. The combined products in (\ref{detc}), (\ref{tc}) are of degree
$N-1$ in each factor, so  (\ref{tc}) is correct to within multiplication by a constant.

To obtain this constant, first look at the contribution to $\det C$ of lowest degree in
$z_{N-1}$ and $z_N$. This is of degree zero in both, so is obtained by setting
$z_{N-1}= z_N =0$.  The last two columns of $C$ then have non-zero entries only in
rows 1 and $N$: 
\be \label{redC}
C = 
\left(  \begin{array}{ccccc}
  .. & .. & .. & 1 & 1 \\
  \# & \# & \# & 0 & 0 \\
\# & \# & \# & ..  & ..  \\
   \# & \# & \#  & 0 & 0 \\
  .. & .. & .. & -1 & 1 \end{array}  \right) \ee
  
  Hence
  \be \det C = 2 (-1)^N \times \;  \overline{C} \ee
  where $\overline{C}$ is the determinant of the $N-2$ by $N-2$ matrix denoted by 
  the $\#$ elements in (\ref{redC}). Hence  $\overline{C}_{j,k} = C_{j+1,k}$,
  $1 \leq j,k \leq N-2$. From (\ref{elemC}), it follows that $\overline{C}$ is the same as
  $C$, with $N$ replaced by $N-2$ and all elements multiplied by $z_1 z_2 
  \cdots z_{N-2}$, i.e
  writing $C$ as $C_N$, 
   \be \det C_N = 2 (-1)^N z_1 z_2 \cdots z_{N-2}   C_{N-2} \period \ee
   
   Iterating, it follows that the term of lowest degree in $z_N, z_{N-1}$, then in 
   $z_{N-2},z_{N-3}$, etc. of $\det C$ is:
   
    \bd 2^{N/2} (z_1 z_2)^{(N-2)/2}  (z_3 z_4)^{(N-4)/2}  \cdots
   (z_{N-3} z_{N-2} )\ed
  if $N$ is even, and  
    \bd-(-2)^{(N+1)/2}z_1^{(N-1)/2}  (z_2 z_3)^{(N-3)/2}  (z_4 z_5)^{(N-5)/2}  \cdots  
    (z_{N-3} z_{N-2} )\ed
     if $N$ is odd.
     
     On the other hand, the corresponding coefficient of the term of lowest degree in the 
  combined products in 
  (\ref{detc}), (\ref{tc}) is 1 except when $N= 3$, mod 4, when it is -1. This minus sign 
  cancels the minus signs in the last equation, leaving the remaining coefficients as 
  $  2^{N/2}$ if $N$ is even, $ 2^{(N+1)/2}$ if $N$ is odd.
  Thus $\tilde{c}$ is as given by (\ref{tc}) and we have proved the identity (\ref{detc}),
   (\ref{tc}) for $\det C$.

\renewcommand{\theequation}{C\arabic{equation}}
\renewcommand{\thesubsection}{C.\arabic{subsection}}

\setcounter{equation}{0}

\section{Calculation of $E$.}

First look at the   factor $\phi \phi'/\eta$ in (\ref{defEG}). From
  (\ref{defphi}), (\ref{defphip}),
\be   \label{phph}
 \phi \phi ' \eq  \zeta^{-2N}
 \prod_{k=1}^N [(1-u z_k)^2-z_k^{2N} (u-z_k)^2 ]\period \ee
 
 Each $z_k$ is a zero of the polynomial $P(z)$ of (\ref{pol}), so 
 \be z_k^{2N} = \frac{(1-tuz_k)(1-uz_k/t) }{ (z_k-tu)(z_k-u/t)} \period \ee
Using the square root of this equation, we can write (\ref{phph}) as
 \be   \label{phph2}
 \phi \phi ' \eq (-1)^n 
 \prod_{k=1}^N \frac{ (1-t)^2(1-u^2) \zeta^{-1} (1-z_k^2)}{[(tu-z_k)(tu-1/z_k)(t/u-z_k)(t/u-1/z_k)]^{1/2} } \ee

 {From} (\ref{pol}), 
 \bd  P(z) = (z^2-1) \prod_{k=1}^N (z-z_k)(z-1/z_k) \comma \ed
 so it is exactly true that
 \bd  \prod_{k=1}^N (t u -z_k)(t u -1/z_k)  = 1-u^2 \ed
\bd  \prod_{k=1}^N (t/u -z_k)(t /u -1/z_k)  = (1-u^2)\, (t/u)^{2N} \ed 
and hence, using (\ref{defmu}),  that
 \be   \label{phph3}
\frac{ \phi \phi ' }{\eta}  \eq \epsilon_N \, i^{-N} \, 
(1-t)^{2N} (1-u^2)^{N-1} \left( \frac{u}{t \zeta}\right) ^N \prod_{k=1}^N (1-z_k)^2  \ee

We now look at the expression $\det Y/{(\det C)}^2$. We have to consider
separately  the cases $N$ even and $N$ odd.

\subsection{$N$ even}

First we focus on the case when $N$ is even and define
\bd n =N/2  \comma \ed
then we  can write (\ref{YC2}) as
\be \label{YC3}
 \frac{\det Y} { (\det C)^2} \eq \frac{ {\cal L} (1-u^2)^{N-1} }{\zeta^{2N}}
  \prod_{j,k =1}^n 
 (z_{2j-1}-z_{2k})^2 (1-z_{2j-1} z_{2k}) ^2 \ee
 
Define polynomials  $P_1(z), P_2(z)$, of degree $N+2, N$, respectively, by
 \ba P_1(z)  & = & (z^2-1) \, \prod_{j=1}^n (z-z_{2j})(z-1/z_{2j})   \comma \nonumber \\
 P_2 (z)  & = &  \prod_{j=1}^n (z-z_{2j-1})(z-1/z_{2j-1})   \period \ea
 Then \be \label{YC4}
 \frac{\det Y} { (\det C)^2} \eq \frac{(1-u^2)^{N-1}}{\zeta^{2N}} \; \prod_{j=1}^n 
 \frac{z_{2j}^N \, (z_{2j}-1)^2  P_1(z_{2j-1})^2}{(z_{2j-1}-1)^2} \ee
and from  (\ref{pol}), 
 \be \label{P1P2}
 P_1(z) \, P_2(z) \eq  P(z) \period  \ee
 
 Consider the functions
 \be \label{deffJ}
 J _{\pm}(z) =  z^N
 \left[ \frac{1-ut/z}{1-t/uz} \right]^{1/2} \pm  \left[ \frac{1- utz}{1-tz/u} \right]^{1/2}  \ee
taking both square roots to be analytic functions of $z$ in an annulus containing the 
unit circle, positive real  when $z $ is real and positive and $t/u < z < u/t$. This 
line segment contains the zeros $z_N,1,  z_{N+1}$ of  $P(z)$.

{From} (\ref{pol}), the zeros of $J_{\pm} (z)$ are also zeros of $P(z)$. By considering the 
limit when $t \rightarrow \infty $, we can see that the zeros of $J_{+}(z)$ are 
the $z_j$ for $j$ odd, i.e. $z_1,z_3, \ldots, z_{N-1}$ and $1/z_1, 1/z_3, \ldots , 1/z_{N-1}$, 
and these are the zeros with \\  $\alpha_j = -1$. 

Thus $J_{+}(z)$ and $P_2(z)$ have precisely the same zeros. Further, if we take $|z|$ to be of 
order 1 and expand the RHS of (\ref{deffJ}) in powers of $t$, then to order  $t^N$, $J_{+}(z)$ 
is a  polynomial in $z$ of degree $N$, with leading term $z^N$. We therefore expect that
if $|z|$ is of order one, then if we neglect terms of order $(t/u)^N$, 
\be \label{P2J}
P_2(z) = J_{+}(z) \period \ee

 Since $J_{+}(z) \, J_{-}(z) \eq P(z)/[(1-uz/t)(1-tz/u)]$, it also follows that
\be \label{P1J} P_1(z) = (1-uz/t)(1-tz/u) J_{-}(z)  \period \ee
Let $z= z_{2j-1}$, then the two terms on the RHS of (\ref{deffJ} )
for $J_{-}(z)$ are equal; taking their geometric mean, we obtain
\be  \label{P1sq}
P_1(z)^2 = -  \, 
\frac{4   z^{N+2} u^2}{t^2 } \, (t/u-z)^{3/2}(t/u-1/z)^{3/2}(ut-z)^{1/2}(ut-1/z)^{1/2}  \ee

Substituting this expression into (\ref{YC4}), we obtain
\be \label{YC5}
 \frac{\det Y} { (\det C)^2} \eq   \frac{(2iu)^{N}  (1-u^2)^{N-1}}{t^N  \, \zeta^N} 
 P_2(t/u)^{3/2} P_2(tu)^{1/2}
 \prod_{j=1}^{n}\frac{z_{2j-1}^{2} (1- z_{2j})^2}{(1-z_{2j-1})^2} \ee
 
 If $|z|<1$, it is still true that (\ref{P2J}) holds to within relative errors of order 
 $(t/u)^N$, provided we only retain the second 
 term in (\ref{deffJ}). Hence in (\ref{YC5}) we can take
 \be \label{valsP2}
 P_2(t/u) =  \left[ \frac{1- t^2}{1-t^2/u^2} \right]^{1/2}  \sep 
  P_2(ut) =  \left[ \frac{1- u^2 t^2}{1-t^2} \right]^{1/2} \period \ee
  
  Now we take the ratio of (\ref{phph3}) to (\ref{YC5}).  Using the definition
  (\ref{defEG}), we obtain
  \be E \eq \frac{(1-t)^{2N} \; \tilde{e}^2}{2^N \, P_2(t/u)^{3/2} \, P_2(tu)^{1/2} }\ee
  where
  \be \tilde{e} \eq \prod_{j=1}^n (1-z_{2j-1}) (1-1/z_{2j-1}) \eq P_2 (1) \period  \ee

  {From} (\ref{deffJ}), (\ref{P2J}), 
  \bd P_2(1) \eq 2 \left( \frac {1-u t}{1-t/u} \right)^{1/2} \comma \ed
  so \be \label{Eeven}
  E \eq \frac{ (1-t)^{2N} (1-ut) \, (1-t^2/u^2)^{3/4} }{2^{N-2} (1-t/u) \,
   (1-u^2 \, t^2)^{1/4} \, (1-t^2)^{1/2}} \; \;  \period \ee

\subsection{$N$ odd}

Taking $N$ to be odd, define the integer
\be n = (N-1)/2  \ee
and the polynomials
\ba \label{defP12odd}
P_1(z)  & = & (z-1) \, \prod_{j=1}^{n+1} (z-z_{2j-1})(z-1/z_{2j-1})   \comma \nonumber \\
 P_2 (z)  & = &  (z+1) \prod_{j=1}^n (z-z_{2j})(z-1/z_{2j})   \period \ea
 Like the $P_1, P_2$ of the $N$ even case, they are of degree $N+2$ and $N$, respectively.
 Eqn. (\ref{P1P2}) is still true and (\ref{P2J}),  (\ref{P1J}) still hold to within terms of relative 
 order  $(t/u)^N$ when $|z| = 1$.
 
 Instead of (\ref{YC4}), we have
 \be \label{YC4B}
 \frac{\det Y} { (\det C)^2} \eq \frac{i \, (1-u^2)^{N-1}}{2 \, \zeta^{2N}} \; \prod_{j=1}^{n+1} 
 z_{2j-1}^{N-1}\,  (1-z_{2j-1}^2)^2 \; 
\prod_{j=1}^{n} 
 \frac{ P_1(z_{2j})^2}{(1-z_{2j})^2} \ee
 
 Since $z_2, z_4, \ldots, z_{2n}$ all lie on the unit circle, we can replace $P1(z)$ 
 in this equation by
 the RHS of (\ref{P1J}). Again the two terms on the RHS of (\ref{deffJ}) are equal, and 
 equation (\ref{P1sq}) still applies.  Using (\ref{defP12odd}), equation (\ref{YC4B}) becomes
 \be \label{YC5odd}
 \frac{\det Y} { (\det C)^2} =  \frac{(-1)^n   i \, 2^{N-2}  \, s \, (u-u^3)^{N-1}\, P_2(t/u)^{3/2} 
 P_2(tu)^{1/2}}{t^{2n}  \, \zeta^{N}
\, (1+t/u)^{3/2} (1+tu)^{1/2}} \; \prod _{j=1}^n \frac{z_{2j}^2}{(1-z_{2j})^2} \ee
 where
  \bd s = \prod_{j=1}^{n+1}\frac{ (1- z_{2j-1}^2)^2}{z_{2j-1}}  \period \ed

{From} (\ref{defP12odd}),  (\ref{P2J}) and   (\ref{P1J}) ,
\bd \prod_{j=1}^{n+1} \frac{(1+z_{2j-1})^2}{z_{2j-1} } \eq -\, \frac{P_1(-1)}{2}  \eq 
\frac{u (1+t/u)^{3/2} (1+ut)^{1/2}}{t} \ed
\bd  \prod_{j=1}^{n} \frac{(1-z_{2j})^4}{z_{2j} ^2} \eq \frac{ P_2(1)^2}{4} \eq \frac{1-ut}{1-t/u}
 \comma \ed
Using these relations and (\ref{valsP2}), taking the ratio of (\ref{phph3}) to (\ref{YC5odd}), 
we obtain
 \be  \label{Eodd}
  E \eq \frac{  (1-t)^{2N} (1-ut) \, (1-t^2/u^2)^{3/4} }{2^{N-2} (1-t/u) \,
   (1-u^2 \, t^2)^{1/4} \, (1-t^2)^{1/2}} \; \;  \comma  \ee
   which is the same as the $N$ even result \ref{Eeven}, even though there are significant
   differences between their calculations.



\begin{thebibliography}{9999}
 
\bibitem{VJ2012} Vernier~E and Jacobsen~J~L 2012 {\myit J. Phys. A: Math. Theor.} 
 {\color{blue} {\mybf 45}} 045003 (41 pages)
 
\bibitem{OB1989} Owczarek~A~L and Baxter~R~J 1989 Surface free energy of
the critical six-vertex model with free boundaries {\myit J. Phys. A:Math. Gen.}
 {\color{blue} {\mybf 22} 1141 -- 1165}
 
\bibitem{RJB2016} Baxter~R~J 2016 Surface and corner free energies of the 
  self-dual square-lattice Potts model {\myit arXiv:}  {\color{blue} 1606.01616}
  
\bibitem{Onsager1944} Onsager ~L  1944 Crystal statistics. I. A 
  two-dimensional model with an order-disorder transition
   { Phys. Rev}  {\color{blue} {\mybf 65} 117--149}
   
\bibitem{Kasteleyn63} Kasteleyn~P~W 1963 Dimer statistics and phase transitions
  {\myit J. Math. Phys.}  {\color{blue} {\mybf 4} 287 -- 293}
  
\bibitem{MPW63} Montroll~E~W, Potts~R~B and Ward~J~C 1963 Correlations and
   spontaneous magnetization of the two-dimensional Ising model
  {\myit J. Math. Phys.}  {\color{blue} {\mybf 4} 308 -- 322}
  
  \bibitem{McCoyWu67}  McCoy~B~M, Wu~T~T 1967 Theory of Toeplitz determinants
  and the spin correlations of the two-dimensional Ising model. IV
  {\myit Phys. Rev.} {\color{blue} {\mybf 162} 436 -- 469}
  
  \bibitem{MCWbook}  McCoy~B~M, Wu~T~T 1973 The two-dimensional Ising model
(Harvard University Press, Cambridge, Mass.,  reprinted Dover 2014, NY)
 
\bibitem{Strog} Stroganov~Yu.~G.  1979 A new calculation method for partition 
functions in some lattice models  { \myit Phys. Lett.}  
 {\color{blue} {\mybf 74A} 116 -- 118}

\bibitem{Bax82} Baxter~R J 1982 The inversion relation for some two-dimensional
 exactly solved models in lattice statistics  {\itt J. Stat. Phys.}   {\color{blue}  {\bff 28} 1--41}
 
\bibitem{Kaufman1949} Kaufman~B  1949 Crystal statistics. II. 
 partition function evaluated by spinor analysis   {\itt Phys. Rev}  
 {\color{blue} {\mybf 76} 1232--1243}

\bibitem{book}  Baxter~R J  1982  Exactly solved models in statistical mechanics,
(Academic, London, re-printed 1989; Dover N.Y. 2007)

\bibitem{Pearce87}  Pearce~P~A 1987
Surface free energies and surface critical behaviour of the ABF models with fixed
boundaries { \itt Phys. Rev. Lett.}   {\color{blue} {\mybf 58} 1502 --1504}

\bibitem{Pearce95}  O'Brien~D~L, Pearce~P~A and Behrend~R~E 1995
Surface free energies and surface critical behaviour of the ABF models with fixed
boundaries { \myit arXiv:}   {\color{blue} 9511081}

\bibitem{Batchelor96}  Batchelor~M~T and Zhou~Y~K 1996
Surface  critical phenomena and scaling in the eight-vertex model
{ \itt Phys. Rev. Lett. }   {\color{blue} {\mybf 76} 14 -- 17}

\bibitem{Pearce97}  O'Brien~D~L and Pearce~P~A  1997
Surface free energies, interfacial tensions and correlation lengths of the 
ABF models { \myit J. Phys. A}   {\color{blue} {\mybf 30} 2353 -- 2366}

\bibitem{Pearcenotes} Pearce~P~A 2012 Bulk, surface and corner free
energies for the anisotropic Potts model {\itt private communication}

\bibitem{CardyPeschel88} Cardy~J~L and Peschel~I 1988 Finite-size 
dependence  of the free energy in two-dimensional critical
systems {\myit  Nucl. Phys. B } {\color{blue} {\mybf 300} 377 -- 392}


\bibitem{Wu2015}  Wu~X and Izmailyan~N 2015 Critical two-dimensional
Ising model with free, fixed  ferromagnetic, fixed  antiferromagnetic and 
double antiferromagnetic boundaries  {\myit
Phys. Rev. E }  {\color{blue} {\mybf 91} 012102, 9pp.}

\bibitem{Abraham95}  Abraham~D~B and LatrŽmolire 1995 {\myit
J. Stat. Phys.  } {\color{blue} {\mybf 81} 539 --559}

\bibitem{Davies97} Davies~B and Peschel~I 1997 {\myit
Ann. Physik  } {\color{blue} {\mybf 6} 187 --214}

\bibitem{GR} Gradshteyn~I~S and Ryzhik~I~M 1965 Table of integrals, series 
and products (Academic, New York \& London)
 
 \bibitem{Yang1952} Yang~C~N 1952 The spontaneous magnetization of a 
 two-dimensional Ising model  {\itt Phys. Rev}  
 {\color{blue} {\mybf 85} 808 --816}
 
 \bibitem{Onsager1971} Onsager ~L  1971 The Ising model in two dimensions
 {\itt in} Mills~R~E, Ascher~E and Jaffee~R~I ``Critical phenomena in alloys, 
 magnets and superconductors",   McGraw-Hill, NY  {\color{blue} pp. 3 --12}
 
 \bibitem{RJB2012} Baxter~R~J 2012 Onsager and Kaufman's calculation of the
 spontaneous magnetization of the Ising model: II {J. Stat. Phys.}
  {\color{blue} {\mybf 149} 1164 -- 1167}
  

 \end{thebibliography}
\end{document}